\documentclass[a4paper,11pt]{article}
\pdfoutput=1 % if your are submitting a pdflatex (i.e. if you have
\usepackage{jcappub} % for details on the use of the package, please
\usepackage{mathtools}
\newcommand{\bq}{\boldsymbol q}

\newcommand{\hq}{\hat{q}}
\newcommand{\hk}{\hat{k}}

\newcommand{\bx}{\boldsymbol x}

\newcommand{\bk}{\textbf{k}}
\newcommand{\bp}{\textbf{p}}
\newcommand{\bL}{\textbf{L}}
\newcommand{\bPsi}{\boldsymbol{\Psi}}
\newcommand{\bs}{\textbf{s}}

\newcommand{\ihmpc}{\,h{\rm Mpc}^{-1}}

\usepackage[normalem]{ulem}

\newcommand{\Tr}[1]{\text{Tr}\left[ #1 \right]}
\newcommand{\TF}[1]{\text{TF}\left\{ #1 \right\}}
\newcommand{\avg}[1]{\left\langle #1 \right\rangle}

\newcommand{\edit}[1]{#1}

\usepackage[T1]{fontenc} % if needed
\usepackage{physics}
\usepackage{aas_macros}

\title{\boldmath A Lagrangian theory for galaxy shape statistics}

\author[a]{Shi-Fan Chen}
\emailAdd{sfschen@ias.edu}
\author[a, b]{Nickolas Kokron}
\emailAdd{kokron@astro.princeton.edu}
\affiliation[a]{School of Natural Sciences, Institute for Advanced Study, 1 Einstein Drive, Princeton, NJ 08540}
\affiliation[b]{Department of Astrophysical Sciences, Princeton University, 4 Ivy Lane, Princeton, NJ 08544}

\abstract{
We formulate the Lagrangian perturbation theory of galaxy intrinsic alignments and compute the resulting auto and cross power spectra of galaxy shapes, densities and matter to 1-loop order. Our model represents a consistent effective-theory description of galaxy shape including the resummation of long-wavelength displacements which damp baryon acoustic oscillations, and includes one linear, three quadratic and two cubic dimensionless bias coefficients at this order, along with counterterms and stochastic contributions whose structure we derive. We compare this Lagrangian model against the three-dimensional helicity spectra of halo shapes measured in N-body simulations by ref.~\cite{Akitsu23} and find excellent agreement on perturbative scales while testing a number of more restrictive bias parametrizations. The calculations presented are immediately relevant to analyses of both cosmic shear surveys and spectroscopic shape measurements, and we make a fast \texttt{FFTLog}-based code \texttt{spinosaurus} publicly available with this publication.
}

\begin{document}
\maketitle
\flushbottom

\section{Introduction}
The last decade of cosmic shear surveys such as the Dark Energy Survey \cite{DESY3}, the Kilo-Degree Survey \cite{KiDS} and the Hyper Suprime Cam Strategic Survey Program \cite{Miyatake23} have demonstrated the unique power of weak gravitational lensing from cosmic shear and its cross correlation with galaxy clustering in probing the growth of cosmic structure. Cosmic shear surveys rely on measurements of the distortion of \emph{shapes} of galaxies, which are subtly sheared and magnified by the distribution of dark matter along the line of sight. In the weak lensing paradigm, we are specifically interested in probing how the quadrupole moment of the projected distribution of galaxy light (also called the galaxy ellipticity) is altered, and the effect of cosmic shear is teased out through correlated measurements of these shapes. The effect of shear is described as an additive contribution to the projected quadrupole moments of galaxies~\cite{Kilbinger_2015}
\begin{equation*}
    g_{ab} (\hat{n}) \approx g^I_{ab} (\hat{n}) + \gamma_{ab} (\hat{n}) + \cdots,
\end{equation*}
where $g^I$ is the assumed intrinsic ellipticity of the galaxy, $\gamma$ is the shear, $\hat{n}$ is the direction on the sky of this field, and $a,b$ are the indices of the matrix which describes the galaxy shape ellipsoid. The shapes of galaxies themselves are not distributed in an uncorrelated fashion, but rather correlated with large-scale structure. These \emph{intrinsic alignments} of galaxies are a key contaminant of the cosmic shear signal~\cite{hirataseljak04,Troxel15} and their contributions to the statistics of shapes must be included in order to infer cosmological parameters in an unbiased fashion~\cite{Krause_2015}. Schematically, the 2-point function between two maps of measured galaxy ellipticities will be
\begin{equation}
    \langle g_{ab}(\hat{n}_1) g_{cd} (\hat{n}_2) \rangle =\langle \gamma \gamma \rangle + \langle g^I \gamma \rangle + \langle \gamma g^I \rangle + \langle g^I g^I \rangle,
\end{equation}
such that the cosmic shear spectrum $\avg{\gamma \gamma}$ receives contamination from cross correlations between cosmic shear and galaxy shape (`GI', `IG') as well as galaxy shape autocorrelations (`II'). Since shear probes the matter density field, such correlations probe the cross-spectrum between shapes and matter
\begin{equation*}
P^{GI} \sim \langle g_{ab} \delta_m \rangle, \quad P^{II} \sim \langle g_{ab} g_{cd} \rangle. 
\end{equation*} 
Studies of cross-correlations between galaxy positions and galaxy shapes (galaxy--galaxy lensing and `3D intrinsic alignment' measurements) will instead probe the `gI' terms
\begin{equation}
P^{gI} \sim \langle \delta_g g_{ab} \rangle.
\end{equation}
While the contributions from cosmic shear are delocalized in space due to the broad support from the lensing efficiency kernel, the contributions from intrinsic alignments depend only on the redshift distribution of the source galaxies whose shapes we are measuring -- their projection kernel is equivalent to that of clustering. This implies that while the challenges of modelling cosmic shear are significant (as they probe deeply nonlinear small scales), models of intrinsic alignments may be formulated purely perturbatively and enjoy comparable success to perturbation theories of the galaxy density used to analyze large-scale structure surveys.\par 
Despite being considered purely a contaminant in cosmic shear surveys, in recent years there has been growing interest in intrinsic alignments as cosmological signals in their own right. In addition to being helpful in understanding the size of intrinsic alignments in galaxies \cite{Mandelbaum06,Singh15}, such measurements can also be used along with standard density statistics to improve cosmological constraints. For example, refs.~\cite{xu2023evidence,Okumura23} recently examined the cross correlation between galaxy shapes and densities in the Sloan Digital Survey to respectively enhance BAO and redshift-space distortions measurements. Beyond these, since shapes are tensors they can potentially probe different primordial physics than densities, for example in the case of anisotropic primordial non-Gaussianity \cite{Schmidt_2015, Akitsu21, Kogai21, Akitsu23b,Kurita23}.\par

There is a long history of perturbative models of galaxy intrinsic alignments (see e.g.~\cite{Troxel15} for a review). Earlier works typically assumed that elliptical galaxies aligned with the local tidal field, giving rise to linear alignment (LA) while spiral galaxies gained their shapes by having their angular momentum align with halos, which in tidal torquing (TT) theory led to a dependence on the \emph{square} of the tidal field~\cite{Catelan_2001}. More recent formulations of galaxy alignments unified these two theories into the TATT (tidal alignment and tidal torquing) model \cite{Blazek_2019}, which was the reference model for intrinsic alignments used in the Dark Energy Survey Year 3 analysis \cite{DESY3}. In particular, ref.~\cite{Schmitz_2018} showed that, accounting for time dependence, the galaxy shape field can be expressed as a linear combination of all rotationally invariant combinations of the Hessian of the gravitational potential, and derived this expansion to second order in perturbation theory. Recently, ref.~\cite{Vlah:2019byq} introduced an effective-theory description of galaxy shapes based on Eulerian perturbation theory. This formulation completes the necessary ingredients for predicting 2-point correlations of intrinsic alignments up to 1-loop order by including third-order bias contributions and effective-theory corrections needed to tame small-scale dependences of the theory, and includes the TATT model as a subset of possible contributions.  This full Eulerian EFT for galaxy shapes has recently been shown to accurately capture the power spectrum of projected halo shapes on perturbative scales \cite{Bakx23}. 

All analytic approaches to modelling the summary statistics of galaxy shapes so far in the literature have relied on Eulerian perturbation theory (EPT), where the shapes of galaxies are expanded perturbatively in terms of their present environment. An alternative method to model the large scale properties of galaxies is Lagrangian perturbation theory (LPT), where e.g. galaxy densities can be written in terms of the initial conditions at their initial (Lagrangian) positions, which are then ``advected'' to their final positions by the flow of dark matter due to gravity. These two approaches can be shown to be perturbatively equivalent in the case of galaxy densities, and indeed can be understood as convenient parametrizations of a more general picture where galaxies respond to their environments along their trajectories across cosmic time (see e.g. \cite{bias_review}). Nonetheless, LPT provides a complementary perspective to structure formation, for example by providing a simple physical picture of the damping of baryon acoustic oscillations by long-wavelength modes and method to resum them~\cite{Matsubara08a, Carlson13, Senatore15, Vlah15b}, as well as simplifying the relation between galaxy bias and dynamical contributions from advection~\cite{Matsubara08b}. LPT is also a natural formalism in which to understand redshift-space distortions, since in it galaxies simply have to be advected by an ``extra'' displacement proportional to their velocities~\cite{Matsubara08a,Carlson13,Chen_2021}. Finally, a method to augment the predictions of LPT with exact dark matter displacements from simulations---termed Hybrid Effective Field Theory---has recently been proposed and shows great promise in future galaxy-galaxy lensing analyses~\cite{Modi_2020, Kokron_2021, Hadzhiyska_2021, DeRose_2023}.

The main purpose of this work is to extend Lagrangian perturbation theory into the realm of galaxy shapes and to derive the associated summary statistics, inspired by the foundational work of ref.~\cite{Vlah:2019byq} in the Eulerian formalism. We are not the first authors to consider this extension---\edit{the Lagrangian bias expansion of galaxy shapes at the field level, up to quadratic order, was developed in ref.~\cite{Taruya21} in the context of super-sample responses and,} while this work was in the early stages of preparation, refs.~\cite{Matsubara22a,Matsubara22b,Matsubara22c} appeared which detailed the formal construction of a Lagrangian theory of shapes within the Lagrangian integrated perturbation theory (iPT)---however to our knowledge we are the first to explicitly construct the necessary bias expansion of shapes up to 1-loop in perturbation theory, including effective-theory corrections, and compute summary statistics within a resummation scheme that respects Galilean invariance. The paper is structured as follows: in \S\ref{sec:lagshapes} we formulate the bias expasion of shapes in terms of Lagrangian displacements, describe their behavior under advection, and relate the resulting contributions to previous results in the literature. This expansion also includes counter-terms and a stochasticity tensor whose properties we derive. In \S\ref{sec:lpt_pk} we review analytic forms and computational techniques for the power spectrum of biased tracers in LPT and extend them to the case of tensor shape fields. Sections \ref{sec:mattershape},  \ref{sec:densshape} and \ref{sec:shapeshape} are the main sections of this work and present calculations of the matter-shape, density-shape and shape-shape correlations in LPT. Finally, we compare our theoretical predictions to N-body data in \S\ref{sec:nbody} before concluding in \S\ref{sec:conclusions}. All of the expressions derived in this paper are numerically implemented and made publicly available in the code \texttt{spinosaurus}\footnote{\href{https://github.com/sfschen/spinosaurus}{https://github.com/sfschen/spinosaurus}}, which builds on top of the framework of \texttt{velocileptors}~\cite{Chen20} to efficiently predict power spectra in LPT.   

\subsection{Notation}
\label{ssec:notation}

Throughout the text we will refer to both Lagrangian $\bq$ and Eulerian $\bx$ coordinates, related via Lagrangian displacements $\bx(\bq,\tau) = \bq + \Psi(\bq,\tau)$. Our main results will be the observed power spectra of galaxy shapes and densities in Eulerian space---we will write these cross spectra between fields $A(\bx)$ and $B(\bx)$, which may be scalars or index-carrying tensors, as
\begin{equation}
    P_{AB}(k) = \avg{ A(\bk) | B(\bk')}' .
\end{equation}
The $'$ denotes the two-point function with a Dirac delta $(2\pi)^3 \delta_D(\bk+\bk')$ removed, and the Fourier transform is with respect to the Eulerian coordinate $\bx$.

In addition to the observed two-point functions in Eulerian space we will deal extensively with \textit{Lagrangian} correlators of the form
\begin{equation}
    \langle O_a(\bq_2)\ O_b(\bq_1) \rangle = \langle O_a(\bq=\bq_2 - \bq_1)\ O_b(\textbf{0}) \rangle = \text{iFT}\{ \langle O_a(\bk) | O_b(\bk') \rangle_{\bq}' \}
\end{equation}
where ``iFT'' refers to the inverse Fourier transform. The subscript $\bq$ refers implies that the Fourier transforms are with respect to $\bq$. 

We will make extensive use of the Legendre basis $P_\ell(\hat{n})$ of tensors, which Fourier transforms into itself:
\begin{align}
    & \int \frac{d^3 \bk}{(2\pi)^3}\ e^{i\bk\cdot\bq} A(k) P_\ell(\hk) = i^\ell P_\ell(\hq) \int \frac{dk\ k^2}{2\pi^2} A(k)\ j_\ell(kq) \nonumber \\
    \label{eqn:tensorlegendrepair}
    & \int d^3\bq\ e^{-i\bk\cdot\bq} A(q) P_\ell(\hq) = 4\pi (-i)^\ell P_\ell(\hk) \ \int dq\ q^2 \ A(q) j_\ell(kq).
\end{align}
These basis tensors are the unique traceless symmetric tensors constructed from e.g., $\hk$, such that for any unit vector $\hat{n}$ we have that $P_\ell(\hk \cdot \hat{n}) =  \hat{n}_{i_1} ... \hat{n}_{i_\ell} P_{\ell, i_1\cdots i_\ell}(\hk)$, from which the above relations can be proven \cite{Guth}. We list them in App.~\ref{app:tensorlegendre}. The Hankel transforms in Equation~\ref{eqn:tensorlegendrepair} can be expressed in terms of \textit{generalized correlation functions} \cite{Schmittfull_2016}
\begin{equation}
    \xi^\ell_n [F](q) = \int_0^\infty \frac{dk}{2\pi^2} k^{2+n} j_\ell (kq) F(k).
\end{equation}
We will often use $\xi^\ell_n$ without the square brackets to refer specifically to Hankel transforms of the linear power spectrum $P_{\rm lin}(k)$. As discussed in~\cite{Schmittfull_2016}, the decomposition of perturbation theory integrals into these particular generalized correlation functions allows for their efficient evaluation as 1D Fast Fourier Transforms using the FFTLog algorithm.\par 
The angular structure of tensor correlators can be significantly simplified by adopting the helicity basis of ref.~\cite{Vlah:2019byq}. This helicity basis is composed of five $\ell=2$ tensor spherical harmonics $Y_{\ell=2,ab}^{(m)} (\boldsymbol{\hk})$ with helicities $m = 0, \pm 1, \pm 2$. These are constructed from tensor products of the Fourier wave-vector $\boldsymbol{\hk}$ and two orthogonal raising/lowering vectors $\boldsymbol{\hat{e}^{\pm}}$ 

\begin{equation}
\label{eqn:tensorsphericalbasis}
Y^{(0)}_{2,ab} = \sqrt{\frac32} \left (\hk_a \hk_b - \frac13 \delta_{ab} \right ), \quad Y^{(\pm 1)}_{2,ab} = \sqrt{\frac12} \left (\hk_a \hat{e}^{\pm}_b + \hk_b \hat{e}^{\pm}_a \right ),\quad Y^{(\pm 2)}_{2,ab} = \hat{e}^{\pm}_a \hat{e}^{\pm}_b.
\end{equation}
We refer the reader to ref.~\cite{Vlah:2019byq} for a detailed derivation of this basis and its properties. 
\section{The Lagrangian Theory of Shapes}
\label{sec:lagshapes}
Within the Lagrangian picture, the galaxy density is modeled by first constructing the initial overdensity of galaxy progenitors in Lagrangian space at their initial positions $\bq$, $\delta_g(\bq)$, then advecting these galaxies to their present day positions $\bx = \bq + \Psi(\bq,\tau)$, where the displacement $\Psi(\bq,\tau)$ defines the trajectory of a fluid element initially at $\bq$. The observed galaxy density is then given by number conservation to be \cite{Matsubara08b}
\begin{equation}
\label{eqn:lptdens}
    1 + \delta_g(\bx, \tau) = \int d^3\bq\ (1 + \delta_g(\bq) ) \ \delta_D(\bx-\bq - \Psi(\bq,\tau)) = (1 + \delta_m(\bx,\tau)) (1 + \delta_g(\bq)).
\end{equation}
The final equality holds in the single-stream regime and derives from the fact that the Jacobian of the transformation between $\bx$ and $\bq$ is given by the matter density. Within this formalism then there is a separation of galaxy clustering due to biasing, encoded in the initial Lagrangian density $\delta_g(\bq)$, and the clustering due to the underlying dynamics, quantified by $\Psi$.

We can similarly model the shape density of a population of galaxies or halos by considering first the shapes of their progenitors. For example, if there are a collection of objects with number density $\delta_g(\bq)$ with moment-of-inertia tensor $I_{ab}(\bq)$ we have that their \textit{shape density} is given by
\begin{equation}
    M_{ab}(\bq) = (1 + \delta_g(\bq)) I_{ab}(\bq),\quad I_{ab}(\bq) = \int d^3\bs\ \rho(\bs, \bq)\ \bs_a \bs_b
\end{equation}
where $\rho(\bs,\bq)$ is the three-dimensional profile of the object centered at $\bq$. As structure forms, each volume element at $\bq$ is advected to $\bq+\Psi(\bq)$, rearranging the distribution of fluid elements withn the object and this, along with number conservation, gives the advected shape density
\begin{equation}
    M_{ab}(\bx) = \int d^3\bq\ \delta_D(\bx - \bq - \Psi(\bq))\ R_{ac}(\bq) R_{db}(\bq) M_{cd}(\bq),\ R_{ab}(\bq) = \delta_{ab} + \nabla_a \Psi_b.
\end{equation}
We see therefore that the shape density is a symmetric rank two tensor  and transforms as such under the transformation $\bq \rightarrow \bq + \Psi(\bq)$, with again an additional multiplication by the matter density at that point. This factor, and the advection integral, implies we always deal with the density-weighted shape in the Lagrangian formalism, as opposed to the shape on its own---it is important to note, however, that the former and not the latter is indeed the physical quantity measured in cosmological surveys, since the spin of an empty region of space is undefined. We will refer to the integral above with and without the factors of $R_{ab}(\bq)$ as \textit{active} and \textit{passive} advection; the latter amounts to the parallel transport of galaxy shapes as they travel along trajectories $\bx(\bq,\tau)$. As we will see, active advection induces contributions to the shapes of galaxies that are entirely degenerate with the Lagrangian bias expansion of $M_{ab}(\bq)$ itself, and as such we will not explicitly include the factors of $R_{ab}$ in most of what follows but rather absorb all contributions to the shape density outside of the $\delta$ function into $\textbf{M}(\bq)$.

It will be useful in the discussion below to decompose these tensors into a scalar trace component and a trace-free component
\begin{equation}
    M_{ab}:\ M = \Tr{M_{ab}}, \TF{M_{ab}} = M_{ab} - \frac13 \delta_{ab} M.
\end{equation}
such that the tensor can also be written as $M_{ab} = \avg{M}( (1 + \delta_M)\ \delta_{ab}/3 + \gamma_{ab})$, with $\gamma_{ab} = \TF{M}_{ab}/\avg{M}$. The former, $\delta_M$, characterizes the relative fluctuation in galaxy size density and is a scalar that follows the familiar bias expansion for e.g. galaxy densities.

\subsection{Bias Expansion}
The Lagrangian shape field $\textbf{M}(\bq)$ is a function of the gravitational environment around which a halo or galaxy forms along its trajectory $\bx(t) = \bq + \Psi(\bq,t)$. Specifically, it can be written in terms of locally observable quantities such as the tidal field $\nabla_i \nabla_j \Phi$ and velocity gradients $\nabla_i v_j$---these can in turn all be written in terms of the Lagrangian shear tensor $L_{ij} = \partial_i \Psi_j$. Significantly, even for initially isotropic protohalos these types of shape terms are generated dynamically, since $R_{ij} = \delta_{ij} + L_{ij}$. This makes sense since gravitational evolution produces sheets and filaments from close-to homogeneous initial conditions even in the Zel'dovich approximation. Since galaxy formation is nonlocal in time---that is, it occurs over the same $\sim H_0^{-1}$ time scale as structure formation itself, it is important to allow contributions at each order $n$ in perturbation theory to have independent coefficients, since their relative effect is a function of which epochs each sample of galaxies is sensitive to.  For example, active advection generates new contributions to the bias expansion by multiplying the Lagrangian shape $M_{ab}$ by $R_{ab}$; however, the coefficients generated in this way should not be taken as fixed since their precise values will depend on, e.g. the time at which a galaxy decouples from large-scale structure following gravitational collapse: the earlier this happens the smaller the effect of higher-order terms. The takeaway, rather is that the dynamical advection of galaxies produces contributions to the shapes of galaxies which are already included in the bias expansion based on $L_{ij}$ evaluated at each order, much like the case for galaxy densities where the relevant expansion is based on the \textit{scalar} functions of $L_{ij}$ \cite{bias_review}.

From the above discussion we can see that a complete bias expansion for galaxy shapes can be built out of the Lagrangian shear tensor of order $n$, denoted by $L^{(n)}_{ij}$, along with other rank-2 functions of it obeying the underlying symmetries. The complete set of such contributions up to third order in perturbation theory is given by:
\begin{align*}
    1^{\rm st} \text{ order: } &L^{(1)}_{ij}\\
    2^{\rm nd} \text{ order: } &L^{(2)}_{ij},\ (\bL^{(1)} \bL^{(1)})_{ij}, \ \Tr{\bL} L^{(1)}_{ij}\\
    3^{\rm rd} \text{ order: }  &L^{(3)}_{ij},\ (\bL^{(2)}\bL^{(1)})_{ij},\ \Tr{\bL^{(1)}} L^{(2)}_{ij},\\
    &(\bL^{(1)}\bL^{(1)}\bL^{(1)})_{ij},\ \Tr{\bL^{(1)}\bL^{(1)}} L^{(1)}_{ij}, \Tr{\bL^{(1)}}^2 L^{(1)}_{ij}, \ \Tr{\bL^{(1)}} (\bL^{(1)}\bL^{(1)})_{ij}.
\end{align*}
Since the shape tensor is symmetric we implictly refer to the symmetrized  versions of these tensors. In the above we have used that $\Tr{\bL^{(n)}}$ can be written in terms of lower-order operators in order to omit certain linearly dependent contributions---note that this degeneracy does not exist at the full tensor level \cite{bias_review,Vlah:2019byq}. In addition, we can construct operators proportional to the Kronecker delta, given by the trace components of the above: $\Tr{O_{ab}} \delta_{ab}$. Up to the Kronecker delta these terms are equivalent to the scalar bias expansion for galaxy densities, yielding one, two and four bias parameters at first, second and third order, respectively~\cite{bias_review}.

The expansion above is a general bias expansion for galaxy shapes, not just the traceless part---in order to better distinguish these we can separate an operator into its trace and trace-free parts: 
\begin{equation}
    O_{ab}(\bq):\ \Tr{O_{ab}(\bq)}, \TF{O_{ab}(\bq)}.
\end{equation}
The trace component is described by the Kronecker delta terms in bias expansion, so performing this split amounts to a linear redefinition of the bias basis. Since we are primarily interested in galaxy shapes in this paper we will implicitly refer only to the trace-free component, and drop the ``TF'' in most cases, unless explicitly noted otherwise. 

In order to make better contact with the literature we note that it is customary to decompose the linear Lagrangian shear as its trace and trace-free components
\begin{equation}
    L^{(1)}_{ij}(\bq) = -\frac{1}{3} \delta(\bq) \delta_{ij} - s_{ij}(\bq)
\end{equation}
as well as to introduce the quadratic operator 
\begin{equation}
    t_{ij} = \Big(\frac{\nabla_i \nabla_j}{\nabla^2} - \frac{1}{3} \delta_{ij} \Big) (\theta^{(2)} - \delta^{(2)}) =  \frac{4}{3} \TF{\bL^{(2)}}_{ij}\footnote{Evaluating the density and velocity potentials in Eulerian perturbation theory leads to the definition of $t_{ij}$ in terms of the quadratic density and tidal field intensities
    \begin{equation*}
    L^{(2)}_{ij} = -\frac{1}{7}\frac{\nabla_i \nabla_j}{\nabla^2} \left [ \big(\delta(\bq)\big)^2 - \frac32 \big(s_{ij}(\bq)\big)^2\right]
    \end{equation*}
    }
\end{equation}
where $\theta$ is the velocity potential. In terms of these operators we can rewrite the above as 
\begin{align*}
    1^{\rm st} \text{ order: } &s_{ij}\\
    2^{\rm nd} \text{ order: } &t_{ij},\ s^2_{ij}, \ \delta s_{ij}\\
    3^{\rm rd} \text{ order: }  &L^{(3)}_{ij},\ (st)_{ij},\ \delta t_{ij},\ s^3_{ij},\ s^2 s_{ij}, \delta^2 s_{ij}, \ \delta s^2_{ij}
\end{align*}
We note that including terms up to second order reproduces the TATT model~\cite{Blazek_2019,Taruya21}, and there are in principle seven new operators that contribute at third order.

\subsection{Cubic Operators in the 1-loop Power Spectrum}
\label{ssec:cubic}
At the one-loop order we consider in this paper, we are interested exclusively in the 2-point correlations of Lagrangian bias operators $O(\bq)$ and displacements $\Psi(\bq)$ up to second order in the linear power spectrum. In this regime contributions due to cubic operators can come only from cross correlation with the linear Lagrangian displacement or density, with one internal loop
\begin{equation}
    O_{ab}(\bk) \rightarrow \left( 3\int_\bp K^O_{ab}(\bk, \bp, -\bp) P_L(p) \right) \delta(\bk) = \left( \frac{k_a k_b}{k^2} - \frac{1}{3} \delta_{ab} \right)\ A_O(k)\ \delta(\bk)
\end{equation}
where $K^O_{ab}$ is the symmetrized kernel for $O$ and $A_O(k)$ is a scalar function of $k$ and form following the last equality follows from symmetry. Note that the Fourier transform here is in \textit{Lagrangian} space. For operators that are built from powers of $\{\delta, s_{ij}\}$ these contributions are entirely degenerate with the linear theory contributions, i.e. $A_O(k)$ is a constant proportional to $\sigma^2$ for these operators, since we can use Wick's theorem to show e.g.
\begin{equation}
    \delta^2 s_{ab}(\bq) \rightarrow \sigma^2 s_{ab}(\bq)
\end{equation}
where the only nonzero contraction is between the two $\delta$'s due to $s_{ab}$ being traceless.\footnote{The other Zel'dovich cubic operators give
\begin{equation}
    s^2 s_{ab} \rightarrow \frac{14}{15} \sigma^2 s_{ab},\ \delta^2 s_{ab} \rightarrow \sigma^2 s_{ab},\ \delta s^2_{ab} \rightarrow 0,\ s^3_{ab} \rightarrow \frac{7}{15} \sigma^2 s_{ab}
\end{equation}
where the right-hand side is the equivalent linear operator that gives the (13) correlation.} For the remaining `non-Zel'dovich operators' their contributions must be equivalent to that of the linear Lagrangian shear $s_{ab}(\bq)$ with the input power spectrum substituted for $A_O(k) P_{\rm lin}(k)$.

In order to compute $A_O(k)$ for the three cubic non-Zel'dovich operators we can compute their cross correlation with $\delta$ in Lagrangian space. Of these we have directly from the correlator of $\Psi^{(3)}$ with $\delta$ \cite{Matsubara08b}
\begin{equation}
    A_{L^{(3)}}(k) = \edit{-} \frac{5}{21} R_1(k),
\end{equation}
where $R_1$ is an integral which is given by the mode-coupling structure of this (13) correlator, defined in ref.~\cite{Matsubara08a}. It is part of a broader family of integrals $R_n$ and $Q_n$ which will be referenced throughout this text and whose expressions are found in refs.~\cite{Matsubara08a, Matsubara08b}. The other two cubic correlators are given by the projections of the tensor (13) integrals
\begin{align*}
    &I^{\delta t}_{ab}(\bk) = -\frac{4}{7}P(k)\int_{\bp} \left( \frac{(\bk-\bp)_a (\bk-\bp)_b}{(\bk - \bp)^2} - \frac13 \delta_{ab} \right) \left ( 1 - \frac{(\bk \cdot \bp)^2}{k^2 p^2} \right ) P(p) \\ 
    &I^{s t}_{ab}(\bk) =-\frac{4}{7}P(k)\int_{\bp} \left ( \frac{\bp_a \bp_c}{p^2} - \frac{\delta_{ac}}{3} \right ) \left( \frac{(\bk-\bp)_c (\bk-\bp)_b}{(\bk - \bp)^2} -\frac13 \delta_{cb} \right) \left ( 1 - \frac{(\bk \cdot \bp)^2}{k^2 p^2} \right ) P(p).
\end{align*}
onto the quadrupole tensor $P_2^{ab}(\hk)$ (Eqn.~\ref{eqn:p2legendre}). Decomposing these projected integrals into generalized correlation functions as in ref.~\cite{Schmittfull_2016} we find
\begin{align*}
    A_{\delta t} (k)  =  -\frac{8}{735} \int &dq q\, \big [ 42 k^2 \xi^0_0 (q) j_0(kq) - 56 k\xi^1_1 (q) j_1 (kq) + 5(-7k^2 \xi^2_0 (q) + 5\xi^2_2 (q))j_2(kq) \\
    &+ 42k \xi^3_1 (q) j_3 (kq) - 18 \xi^4_2(q) j_4(kq) \big ] + \frac{8}{105} \sigma^2\\ 
    A_{st}(k) = -\frac{8}{2205} \int& dq q\, \big [ -38 k \xi^1_1(k) j_1(kq) + 25(k^2 \xi^2_0(q) + \xi^2_2) j_2(kq) + 14 k \xi^3_1(q) j_3(kq)\\
    &- 18 (k^2 \xi^4_0(q) + \xi^4_2) j_4(kq) + 10 k \xi^5_1(q) j_5(kq) \big ] + \frac{8}{315} \sigma^2.
\end{align*}
In the above we have explicitly included the zero-lag piece proportional to $\sigma^2$. In general these contributions can be removed by appropriate defining a ``normal-ordered'' operator, equivalent to a linear combination of the original operator with lower-order ones, with 
\begin{equation*}
:O^{(3)}_{ab}: = O^{(3)}_{ab} - A_O(k\rightarrow0)\ s_{ab},
\end{equation*}
which we will implicitly assume throughout the rest of this work.

Finally, as ref.~\cite{Vlah:2019byq} first discussed, various degeneracies between the bias parameters exist at one-loop level for traceless galaxy shapes; some algebra shows that
\begin{equation*}
    A_{st} - \frac{1}{6} A_{\delta t} \edit{+} \frac{3}{5} A_{L^{(3)}} = \frac{1}{7} \int_\bp \left( \frac{11}{45} -\frac43 \mu^2 +\mu^4 \right) P(p) = 0.
\end{equation*}
This implies we only need to introduce two bias parameters at this order; since $A_{L^{(3)}}$ is shared with the density bias case we will adopt it and $\delta t_{ab}$ as our conventional choice. This is slightly different from the case of cubic operators in the perturbation theory of scalar fields, where only one non-degenerate transfer function $\sigma^2(k)$ and effective cubic parameter `$\tilde{b}_3$' are introduced \cite{McDonald09}. \footnote{\edit{Specifically, we note that the effective cubic parameter in refs.~\cite{Chen20,Chen_2021} is defined using the operator $O_3 = \Tr{st}$ such that its cubic kernel is $R_{b_3} = -\frac{12}{5} A_{L^{(3)}}$.}}

\subsection{Effective Theory Considerations}

In order to complete the list of perturbative contributions to galaxy shapes we need to include dimensionful coefficients reflecting that our perturbation theory is an effective one. The first type are the so-called counterterms or derivative biases, which in the bias expansion reflect the nonlocality of astrophysical processes, i.e. that the formation and shape of a galaxy depends not on its environment not at a point but in a neighborhood around it with size given roughly by its Lagrangian radius $R_h$. This gives rises to contributions proportional to higher derivatives such as
\begin{equation}
    \delta_g(\bq) \ni \alpha_g \nabla^2 \delta(\bq), \quad M_{ab}(\bq) \ni \alpha_M \nabla^2 s_{ab}(\bq), \quad \alpha_{g,M} \sim \mathcal{O}(R_h^2).
\end{equation}
Similar considerations require additional contributions to the dynamics, i.e. displacements \cite{Vlah15b}. At 1-loop order, where we will work, all of these contributions lead to corrections to the power spectrum $k^2 P(k)$, with a tensor structure given either by the scalar $\delta$ (density bias and dynamics) or trace-free tensor $s_{ab}$ (shape bias).

Beyond the counterterms we also need to account for stochastic contributions, which quantify the effect of short wavelength physics which we do not explicitly model. We will be most concerned with contributions uncorrelated with long wavelengths, 
\begin{equation}
    \delta_g \ni \epsilon(\bq), \quad M_{ab} \ni \epsilon_{ab}(\bq).
\end{equation}
Since these encapsulate small-scale physics their correlation will be localized to $|\bq_1 - \bq_2| \lesssim R_h$, where $R_h$ is the halo radius. This localization highly restricts the form these stochastic terms can contribute to the power spectrum---we will return to this topic in \S\ref{sec:stochastic}.\footnote{Beyond these contributions the stochastic contribution to active advection can also help us explain the smallness of the intrinsic alignment signal---whereas galaxy bias parameters tend to be order one, galaxy shape biases tend to be of order a few percent when they are detected at all.

Let small-scale astrophysics contribute some additional rotational ``kick'' to the active advection so that the shape goes to
\begin{equation}
    M (1 + \gamma^{\rm PT}) \rightarrow \textbf{R}_{\rm stoch}^T(\bx) M (1 + \gamma^{\rm PT} ) \textbf{R}_{\rm stoch}(\bx)
\end{equation}
where $M$ and $\gamma$ are the perturbative size and shape of the galaxy. \edit{At first sight it may seem like since both the trace $M$ and $\gamma$ are multiplied by the same matrix $\mathbf{R}$ any change in the shape would be normalized out by considering the corresponding change in $M$; however, the trace and trace-free parts do not transform equivalently under coordinate transformations, so the change induced by this additional rotation cannot in general be accounted for by the normalization $M$.} 

Since the stochastic terms do not correlate with the large-scale structure by construction we can compute their average ahead of time, i.e. 
\begin{equation}
    M (1 + \gamma^{\rm PT})_{ij} \rightarrow M \langle \textbf{R}_{ia,\rm stoch} \textbf{R}_{ja, \rm stoch} \rangle + M \langle \textbf{R}_{ia,\rm stoch} \textbf{R}_{jb, \rm stoch} \rangle \gamma^{\rm PT}_{ab}
\end{equation}
Let us take as a\edit{ toy }example the case where $\textbf{R}$ imposes a random rotation \edit{$\vec{\Omega} = \omega\  \hat{\omega}$} about axis \edit{$\hat{\omega}$ by a fixed angle $\omega$}. \edit{The trace component $M$ will not be changed by such a transformation for any $\Omega$}, but it is straightforward to show that, \edit{averaging over orientations $\hat{\omega}$,} we have
\begin{equation}
    \gamma \rightarrow \frac{1}{5} \big(1 + 2\cos\omega + 2\cos(2\omega) \big)\ \gamma.
\end{equation}
\edit{The function multiplying $\gamma$ becomes equal to $1$ when $\omega = 0$, but is generally less than one, implying that rotations induced by small-scale physics that are not aligned with large-scale shape fields, i.e. have some random orientation $\hat{\omega}$ that we can average over, will tend to wash out the traceless shape of galaxies. More generally, we can see by symmetry that the second moment of the stochastic transformations must take the form
\begin{equation}
    \langle \textbf{R}_{ia,\rm stoch} \textbf{R}_{jb, \rm stoch} \rangle = \frac{r_0}{3} \delta_{ij} \delta_{ab} + \frac{r_2}{2} ( \delta_{ia} \delta_{jb} + \delta_{ib} \delta_{jab} )
\end{equation}
where $r_{0,2}$ are two independent components. Importantly, the trace component $M$ gets multiplied by $r_0 + r_2$ while the symmetric trace-free shape $\gamma$ only gets multiplied by $r_2$, showing that the sizes and shapes are galaxies are affected independently. Note that rotations satisfy $r_0 + r_2 = 1$ such that only the latter component is affected.}}

\section{Power Spectra in Lagrangian Perturbation Theory}
\label{sec:lpt_pk}
The expressions we have written down so far have been in terms of the full three-dimensional components of galaxy shape tensors $M_{ab}$. In order to evaluate the $n$-point correlation functions of these tensors, and their cross correlations with galaxy density, however, it will be more convenient to project the tensors onto an irreducible basis under rotations, and to compute the spectra of these independent components. Such a spherical harmonic basis was proposed in ref.~\cite{Vlah:2019byq}, and we will adopt it in this paper; in particular a traceless symmmetric tensor like the galaxy shape can be decomposed as a sum of helicity modes $m$ of $\ell=2$ tensor spherical harmonics
\begin{equation}
    M_{ab}(\bk) = \sum_{m=-2}^2 M_m(\bk) Y_{2,ab}^m(\hk). \nonumber
\end{equation}
Enforcing rotational symmetry about $\bk$ requires that the only nonzero correlations are between components of the same helicity or, as a special case, the helicity-zero component with a density.

Within this basis it is easy to see that there is only one nonzero component of the galaxy density-shape power spectrum, which can be written as a quadrupole operator and a scalar spectrum
\begin{equation}
\label{eqn:densshape_corr}
    \left\langle \delta_g(\bk) | M_{ab}(\bk')  \right\rangle' = \frac{3}{2} \left( \hk_a \hk_b - \frac{1}{3} \delta_{ab} \right) P_{gI}(k).
\end{equation}
For the shape-shape autocorrelations, there will tend to be three components due to the fact that the three helicity modes do not correlate \cite{Vlah:2019byq}\footnote{This projection is motivated by properties of the spherical tensor basis derived in ref.~\cite{Vlah:2019byq}. The general power spectrum of two shape fields can be written as
\begin{equation}
    \avg{M_{ab}(\bk) M_{cd}(\bk')} = \sum_{m=-2}^2 Y_{2,ab}^m(\hk) Y_{2,cd}^m(-\hk) P_{2m}(k) =  \sum_{m=-2}^2 (-1)^m Y_{2,ab}^m(\hk) Y_{2,cd}^{-m}(\hk) P_{2m}(k).
\end{equation}
In order to extract the $P_{2m}$ we can dot each spherical tensor with its complex conjugate, given by $Y^{m\ast}_2(\hk) = (-1)^m Y_2^{-m}(\hk)$, where the two factors of $(-1)^m$ from the two conjugated $Y_2^m$'s cancel. Noting that $P_{2m} = P_{2(-m)}$ due to parity then yields the symmetried projection and factor of $(-1)^m/2$.
}
\begin{equation}
    P_{2m}(k) = (-1)^m \frac12 \left( Y_2^m(\hk) Y_2^{-m}(\hk) + Y_2^{-m}(\hk) Y_2^{m}(\hk) \right)_{abcd} \left\langle M_{ab}(\bk) | M_{cd}(\bk')  \right\rangle'.
    \label{eqn:helicity_components}
\end{equation}
This decomposition implies the angular structure of shape--shape correlators is simpler than naively anticipated given the number of tensor components -- and a reduced number of spectra need to be computed in order to describe the two-point statistics of shapes and their cross-correlations with density. For the rest of this paper we will compute the quadrupole component density--shape correlations as in \ref{eqn:densshape_corr} and the three components of the shape--shape correlations as shown above.

\subsection{The Generating Function}

We now describe the computing of power spectra of biased tracers in Lagrangian Perturbation Theory. From Equation~\ref{eqn:lptdens}, the power spectrum between two fields $\alpha,\beta$ in LPT is given by \cite{Matsubara08b,Carlson13}
\begin{equation}
    P(\bk) = \int d^3\bq\ e^{i\bk\cdot\bq} \langle e^{i\bk\cdot\Delta(\bq_1,\bq_2)} F_\alpha(\bq_1) F_\beta(\bq_2) \rangle_{\bq=\bq_2-\bq_1}
    \label{eqn:lpt_spectra}
\end{equation}
where $\Delta = \Psi(\bq_2) - \Psi(\bq_1)$. The functions $F_{\alpha,\beta}(\bq)$ denote the unadvected values of these fields, i.e. the proto-halo density $1 + \delta_g(\bq)$ for densities and the progenitor shape distribution $(1 + \delta_g(\bq)) I_{ab}(\bq)$ for shapes. We have dropped a delta function at $\bk=0$ that would appear in the case of density-density statistics and suppressed tensor indices that would appear in the case of spin statistics.

The $F_{\alpha,\beta}$ are functions of the local environment that can be perturbatively expanded. In order to derive expressions for the two-point function of scalar or tensor biased tracers in LPT we make use of the generating function\footnote{An alternative treatment would be to treat the sources $\alpha_i$ as $\bq$-dependent functions -- that is the field and coefficients would be integrated as $\int d^3 \bq\ \alpha_i (\bq) O_i (\bq)$ and functional derivatives taken to isolate particular bias operators. These two approaches are equivalent, but the functional derivative approach is more natural when dealing with higher-order statistics where many more $\alpha_n$ would be involved.}
\begin{equation}
\label{eqn:genfunc}
M(\boldsymbol{\alpha}) = \avg{ \exp{ i\bk\cdot\Delta + \left[\sum_a \alpha_1^a O^a(\bq_1) + \alpha_2^a O^a(\bq_2) \right] } }.
\end{equation}
which can be evaluated via the cumulant expansion. For example, the $\boldsymbol{\alpha}=0$ piece is given by
\begin{equation*}
    M(\mathbf{0}) = \exp \left[ -\frac12 k_i k_j A_{ij}(\bq) - i\frac16 k_i k_j k_k W_{ijk}(\bq) + \cdots \right], \quad A_{ij}(\bq) = \langle \Delta_i \Delta_j \rangle, W_{ijk}(\bq) = \langle \Delta_i \Delta_j \Delta_k \rangle.
\end{equation*}
Ignoring the self-pairs we have
\begin{align}
\frac{M(\boldsymbol{\alpha})}{M(\mathbf{0})} =  & \exp{ \Big[ \sum_a (\alpha_1^a + \alpha_2^a) ik_i \langle \Delta_i O_1^a \rangle + \sum_{a,b}\alpha_1^a \alpha_2^b \langle O_1^a O_2^b \rangle  \Big] } \nonumber \\
& \exp{ \Big[ -\frac{1}{2} \sum_a  (\alpha_1^a + \alpha_2^a) k_i k_j \langle \Delta_i \Delta_j O_1^a \rangle_c  + \sum_{a,b} i \alpha_1^a \alpha_2^b  k_i \langle \Delta_i O_1^a O_2^b \rangle_c + ...\ \Big] }
\label{eqn:generating_function}
\end{align}
where we have separated the exponentials by cumulant, truncating at cubic order. We have also dropped any terms corresponding to more than one $\alpha$ from any given point, or that is more than quadratic in $\alpha$, since these will not appear in the 2-point function.  The correlators in the above equation are invariant under inversion $\bq \rightarrow -\bq$, which takes $\Psi(\bq) \rightarrow - \Psi(-\bq)$ and thus, along with a translation, gives $\Delta \rightarrow \Delta$, $O_1 \rightarrow O_2$.

Given the generating function above, matter--tracer and tracer--tracer correlation functions are computed by taking derivatives with respect to $\alpha_i$ and then setting these `sources' to zero. Carrying out these derivatives the power spectra in question are given by
\begin{equation}
    P_{mt}(\bk) = \sum_a b_a \int d^3\bq\ e^{i\bk\cdot\bq} \left. \frac{\partial M}{\partial \alpha_2^a} \right | _{\boldsymbol{\alpha}=0}, \quad P_{tt} = \sum_{a,b} b_a b_b \int d^3\bq\ e^{i\bk\cdot\bq} \left. \frac{\partial^2 M}{\partial \alpha_1^a \partial \alpha_2^b} \right |_{\boldsymbol{\alpha}=0}
    \label{eqn:power_spectra_master}
\end{equation}
where we have
\begin{equation}
\left. \frac{\partial M}{\partial \alpha_2^a} \right |_{\boldsymbol{\alpha}=0} = M(\mathbf{0}) \Big( i k_i \langle \Delta_i O_2^a \rangle - \frac{1}{2} k_i k_j \langle \Delta_i \Delta_j O_2^a \rangle + ... \Big) \nonumber
\end{equation}
\begin{equation}
\mathclap{\left. \frac{\partial^2 M}{\partial \alpha_1^a \partial \alpha_2^b} \right |_{\boldsymbol{\alpha}=0}\quad\quad} = M(\mathbf{0}) \Big( -k_i k_j \langle \Delta_i O_1^a \rangle\langle \Delta_j O_2^b \rangle + \langle O_1^a O_2^b \rangle + ik_i \langle \Delta_i O_1^a O_2^b \rangle_c + ... \Big).
\end{equation}
These expressions are `master formulae' for the power spectra of biased tracers in LPT. Each of the correlators $\langle \Delta_i O \rangle$, $\langle \Delta_i \Delta_j O \rangle$ and $\langle O_1 O_2 \rangle$ are then evaluated in pertubation theory up to fourth--order in fields. The complete order-by-order structure of these correlators which contribute to biased tracer spectra are enumerated in Appendix~\ref{app:cumulant_contributions}. 

\subsection{IR Resummation and the Eulerian Limit}

One of the key features of Lagrangian perturbation theory is that it allows for a straightforward treatment of long-wavelength linear displacements, which lead to large nonlinearities such as the damping of baryon acoustic oscillations (BAO), and hence must be resummed. Within the generating function, these effects are captured by the linear contribution to $M(\boldsymbol{\alpha})$, i.e. $A^{\rm lin}_{ij}(\bq) = \langle \Delta^{(1)}_i \Delta^{(1)}_j \rangle$. In particular, it is desirable to split the linear displacements above and below some exponential cutoff $S(k) = e^{-(k/k_{\rm IR})^2/2}$ and resum the long-wavelength piece while expanding the short-wavelength piece to 1-loop order such that
\begin{equation*}
    M^{\rm 1loop}(\boldsymbol{0}) = \exp\left[-\frac{1}{2} k_i k_j A^<_{ij}(\bq) \right] \left( 1 - \frac{1}{2} k_i k_j (A_{ij}^>  + A_{ij}^{\rm loop}) - \frac i 6 k_i k_j k_k W_{ijk} + \frac18 k_i k_j k_k k_l A_{ij}^> A_{kl}^>  \right)
\end{equation*}
and similarly 
\begin{equation}
M^{\rm 1loop}(\boldsymbol{\alpha}) = \exp\left[-\frac{1}{2} k_i k_j A^<_{ij}(\bq) \right] \left( 1 + M^{\rm linear}(\boldsymbol{\alpha}) + M^{\rm loop}(\boldsymbol{\alpha}) \right).
\label{eqn:M_alpha}
\end{equation}
\edit{In the above $A_{ij}^{<}$ and $A_{ij}^{>}$ denote the linear contributions to $A_{ij}$ due to modes below the exponential cutoff and above it, respectively as defined in refs.~\cite{Chen20, Chen_2021}, and} the linear and loop terms in parentheses have their contributions from long displacements resummed in the exponential. From the above expressions it is also possible to extract the un-IR-resummed Eulerian predictions for shape statistics; these are simply equivalent to setting $k_{\rm IR} = 0$, i.e. fully expanding the exponentiating exponents to 1-loop order \cite{Chen20}. We can also extract a different resummation based on splitting the linear power spectrum into its ``wiggle'' and ``no-wiggle'' components by noting that any Lagrangian correlators with a peak at the BAO scale will pick out the exponentiated piece in Equation~\ref{eqn:M_alpha} in the $\bq$ integral in Equation~\ref{eqn:lpt_spectra} \cite{Vlah16}. Performing this saddle-point approximation yields the IR-resummed form for shapes derived in Equation 4.24 in ref.~\cite{Vlah:2019byq}.

\subsection{Angular Structure of Lagrangian Correlators}
\label{ssec:angular}
Having laid out the general structure of tensor power spectra in Lagrangian perturbation theory, much of the remaining work of this paper will be in calculating the correlators between bias operators and pairwise displacements that compose the generating function in Equation ~\ref{eqn:generating_function}. Before we proceed it will thus be helpful to describe the general structure of these correlators. We are interested in the Lagrangian space correlators of the form
\begin{equation}
    \avg{\Delta_{i_1} ... \Delta_{i_n} O_{ab}(\bq_1) O_{cd}(\bq_2) }, \quad \avg{\Delta_{i_1} ... \Delta_{i_n} O(\bq_1) O_{ab}(\bq_2) }
\end{equation}
where, in the case of cross correlations with matter we have $O=1$. By symmetry, any such correlators must be expressible as functions of $q = |\bq|$ multiplied by products of Kronecker $\delta$'s and the unit vector $\hq$. However, since perturbation-theory solutions for these operators are typically written in Fourier space it will be necessary to first compute the Fourier transforms of these correlators before converting them into configuration (Lagrangian) space.

As a first example let us consider the linear density-shear correlator $E_{ab}(\bq)$. Taking its Fourier transform we have that 
\begin{equation}
    E_{ab}(\bq) = \avg{\delta(\bq_1) s_{ab}(\bq_2)} = \text{FT}\left\{  \left(\frac{k_a k_b}{k^2} - \frac13 \delta_{ab} \right) P_{\rm lin}(k) \right\} = - \xi^2_0(q) (\hq_a \hq_b - \frac13 \delta_{ab})
\end{equation}
where we have used Equation~\ref{eqn:tensorlegendrepair} in the final step. This example is particularly straightforward because it and its Fourier transform can be written in terms of a single Legendre basis tensor. Typically this will not be the case and the Fourier-space spectra have to be suitably decomposed.

The most general way to write down these correlators in Fourier space is in the helicity basis defined in ref.~\cite{Vlah:2019byq}. Since the properties of this basis are extensively documented in that work we will only summarize the relevant pieces here. Briefly, the traceless symmetric tensors of rank 2, i.e. shapes, can be described as the spin 2 component of $1 \otimes 1 = 2 \oplus 1 \oplus 0$, which is spanned by $\ket{2,m}$ with $m = 0, \pm1, \pm2$, represented by the spherical harmonic tensors $Y_{\ell=2,m}$ defined in Equation~\ref{eqn:tensorsphericalbasis}. Scalars, i.e. densities, are described by the spin 0 part, $Y_{0,0} = 1$. For correlators involving purely shape and density operators, i.e. which do not contain factors of the displacement $\Psi_i$, symmetry under rotations about $\bk$ requires that only projections of the same helicity $m$ are nonzero, such that the correlators can be decomposed into sums of the products $Y^m_\ell(\hk) Y^{-m}_{\ell'}(\hk)$. Importantly, these products themselves also have cancelling azimuthal dependence such that they can re-projected into the Legendre basis \cite{Sakurai} and Fourier-transformed as in Equation~\ref{eqn:tensorlegendrepair}.

For correlators involving one or more displacements the picture is similar, except now we must deal with operators of the form
\begin{equation}
    k_{i_1} ... k_{i_m} \left( \Psi_{i_1} ... \Psi_{i_m} O_{ab} \right)(\bq).
\end{equation}
where we have kept the powers of the wave-vector $\bk$ accompanying each displacement in LPT explicit, in the case of shapes. The case for density operators is similar. The tensor operator in parentheses is now a rank $m + 2$ tensor and as such must now be expressed in the subspace of rank $m+2$ tensors $1 \otimes \cdots \otimes 1$ where the first $m$ indices are symmetric and the last two are symmetric and traceless. When long-wavelength linear displacements aren't resummed, i.e. in the Eulerian theory, we can further express the dot product with the wavevector in front as dot products with the angular moment eigenstate $\ket{1,0}$, such that the relevant \textit{projected} operators which contribute to the shape statistics with helicity $m$ are restricted to the $(\bigotimes^n \ket{1,0} ) \otimes \ket{\ell,m}$ components of the product operator in parentheses. In LPT, however, the distinction between Lagrangian and Eulerian coordinates and wavevectors due to resummed displacements makes this observation less useful. Furthermore, as we  discuss in \S\ref{sec:conclusions} it is of general interest to evaluate the full tensorial structure of displacement-shape correlators, for example to compute the effect of redshift-space distortions, rather than restricting to the components selected out by dotting with the wavevector.  This structure however can be slightly simplified: since any two-point correlators, with or without displacements, must still have azimuthal symmetry about $\bk$, they will nonetheless be forced to have total helicity zero and be expressible in terms of the Legendre tensor basis.

Let us now explicitly describe how to evaluate these correlators in terms of Legendre basis tensors. As an example let us take correlators of the form
\begin{equation}
    \avg{O_{ab}(\bk) |  \Psi_i(\bk')}' = F_3(k) \mathcal{Q}^3_{iab}(\hk) + F_1(k) \mathcal{Q}^1_{iab}(\hk).
    \label{eqn:Qiab}
\end{equation}
By the above arguments this correlator must be expressible in terms of the Legendre basis and, due to parity and rank, must be expressible solely in terms of $P_{1,3}$. The tensor $P_3$ can immediately be seen to be a possible component of the correlator as it is both traceless and symmetric under all indices; we thus define $\mathcal{Q}_{iab}(\hk) = P_{3,iab}$. Since $P_{1,a}(\hk) = \hk_a$ is rank 1 it must be multiplied by factors of the Kronecker $\delta$. The two possible combinations are
\begin{equation*}
    P_{1,i}(\hk) \delta_{ab}, \quad P_{1,(a}(\hk) \delta_{b)i}
\end{equation*}
where we have used that the indices $a,b$ must be symmetric. However, since this basis must additionally be traceless in $a,b$ we have that the only viable combination is
\begin{equation}
    \mathcal{Q}^1_{iab}(\hk) = P_{1,i} \delta_{ab} - \frac32 P_{1,(a} \delta_{b)i}
\end{equation}
leading to the two components shown above. Since $\mathcal{Q}^{1,3}$ are written in terms of $P_\ell$ this expression can then be straightforwardly Fourier transformed to yield
\begin{equation}
    \avg{O_{ab}(\bq_2) \Psi_i(\bq_1)} = -i \xi^3_0\left[F_3 \right](q) \mathcal{Q}^3_{iab}(\hq) + i \xi^1_0 \left[F_1\right](q) \mathcal{Q}^1_{iab}(\hq).
\end{equation}
In general, we can generate an appropriate Legendre basis for two-point correlators with $n$ indices as follows:
\begin{enumerate}
    \item For a given Legendre tensor $P_\ell(\hk)$ of the same parity, form all tensor products with Kronecker $\delta$'s with combined rank $n$.
    \item Symmetrize each product under the pairwise and exchange symmetries and keep the linearly independent pieces.
    \item Solve for the linear combinations of these pieces that are trace-free in the shape indices to generate the full basis.
\end{enumerate}
We will make extensive use of this construction throughout the rest of the paper and denote the components as $\mathcal{Q}^\ell_{ijk... abc...}$ where $ijk$ and $abc$ refer to displacements and shape indices as usual. Note that the basis elements with different $\ell$ will be orthogonal, that is to say $\mathcal{Q}^\ell_{ijk...abc...} \mathcal{Q}^{\ell'}_{ijk...abc...} = 0$ for $\ell \neq \ell'$ due to $P_\ell$ being traceless.

\subsection{Angular Structure of Stochastic Terms}
\label{sec:stochastic}
We can also infer the shape of stochastic contributions to shape-shape and density-shape power spectra using the mathematical tools developed in this section. The shape of these contributions follows from the fact that their correlations are limited in support to small, non-perturbative scales. In the case of the density this implies $\avg{ \epsilon(\bq_1) \epsilon(\bq_2) } = A f(q/R_h)$ for some unit-normalized function $f$. The stochastic power spectrum is then
\begin{equation}
    \avg{\epsilon(\bk) | \epsilon(\bk')}' = A \int d^3 \bq\ e^{i\bk \cdot \bq} \ f\left(\frac{q}{R_h}\right) = A R_h^3 \left(1  - \frac12 \sigma^2 R_h^2 k^2 + \mathcal{O}(R_h^4 k^4) \right)
\end{equation}
where $\sigma^2 \sim 1$ is the dimensionelss second moment of $f$. Equivalently the form above can be derived by approximating the distribution $f$ as a series of $\delta$ functions. For galaxies, $\epsilon$ is qualitatively well-approximated by a Poisson-sampled random field~\cite{Baldauf_2013}, in which case the quantity $A R_h^3$ is given by the inverse number density $\bar{n}_g^{-1}$. If the galaxies are sufficiently dense then the scale-dependent corrections tend to be small and can often be neglected.

For galaxy and halo shapes the situation is similar, with the caveats that (1) the function $f_{abcd}$ is now a tensor and (2) the so-called ``shape noise'' of galaxies is typically somewhat larger than the large-scale clustering signal itself on scales of interest. Accounting for the symmetries of galaxy shapes the leading-order contribution to the shape noise is given by \cite{Vlah:2019byq}
\begin{equation}
     \avg{\epsilon_{ab}(\bk) | \epsilon_{cd}(\bk')}' = B R_h^3 \left( \delta_{ac} \delta_{bd} + \delta_{ad} \delta_{bc} - \frac23 \delta_{ab}\delta_{cd} \right)
\end{equation}
i.e. there is only one linearly independent contribution that does not depend on the vector $\bk$. This is equivalent to constructing $\mathcal{Q}^0_{abcd}$ with the symmetries of this correlator.

At next to leading order in the expansion around the zero-lag limit of $q\to0$ we have the scale-dependent contribution 
\begin{equation*}
    -\frac12 k_i k_j \int d^3\bq\ q_i q_j \ f_{abcd}\left(\frac{\bq}{R_h}\right) \sim - \frac{R_h^5}{2} \times k_i k_j \left( \delta \delta \delta \right)_{ijabcd}
\end{equation*}
where the factor of $(\delta\delta\delta)$ denotes the angular structure of the second moment of $f_{abcd}$. Again using the symmetries of the problem it is not hard to see that there are only two independent combinations that involve at most two powers of the unit wavevector $\hk$, equivalent to $k^2 \mathcal{Q}^0_{abcd}$ and $k^2 \mathcal{Q}^2_{abcd}$:
\begin{equation}
    k^2 \left( \delta_{ac} \delta_{bd} + \delta_{ad} \delta_{bc} - \frac23 \delta_{ab}\delta_{cd} \right), \quad k^2 \left( Y_{2}^1(\hk) Y_{2}^{-1}(\hk) + 4 Y_{2}^2(\hk) Y_{2}^{-2}(\hk) \right)_{\rm sym}.
\end{equation}
This implies that the $k^2$ stochastic contributions to the three helicity spectra are not independent but rather described by two free parameters. Similar logic implies that the cross-stochasticity can be shown to have no scale-independent piece
\begin{equation}
    \avg{\epsilon(\bk) | \epsilon_{ab}(\bk')} = C R_h^5 k^2 \left(\hk_a \hk_b - \frac13 \delta_{ab} \right) + ...
\end{equation}
such that the leading order contribution is a quadrupole proportional to $k^2$.

\subsection{Numerical Evaluation}

From the above discussion it is clear that the Lagrangian correlators in Equation~\ref{eqn:generating_function} can be written in terms of Kronecker $\delta$'s and powers of $\hq_i$, multiplied by scalar functions of the separation $q$. When dotted with the appropriate wave-vectors and spherical harmonics to obtain the components in Equation~\ref{eqn:helicity_components} this therefore yields the form~\cite{Vlah15b,Vlah_2016,Chen20}
\begin{align*}
    \int d^3\bq\ &e^{i\bk\cdot\bq - \frac12 k_i k_j A^<_{ij}(\bq)}\ c_n(q) \mu^{2n(+1)} \nonumber\\
    &= i^{0(+1)}4\pi \int dq\ q^2\ e^{-\frac12 k^2 (X^<(q) + Y^<(q))} \sum_\ell f^{2n}_\ell(q) \left( \frac{kY^<(q)}{q}\right)^\ell c_n(q) j_{\ell(+1)}(kq).
\end{align*}
Here $c_n$ is a scalar function of $q$, $\mu = \hq \cdot \hk$ and we have defined the kernel using confluent hypergeometric functions of the second kind
\begin{equation}
    f^{2n}_\ell(q) = \left( \frac{2}{k^2 Y^<}\right)^n U \left(-n, \ell - n + 1, \frac{k^2 Y^<}{2}\right).
\end{equation}
These integrals then take the form of Hankel transforms and can be rapidly performed using the FFTLog algorithm~\cite{fftlog}.

\section{Matter-Shape Correlators at 1-loop}
\label{sec:mattershape}

We now proceed to use the bias expansion and mathematical framework described in \S\ref{sec:lagshapes} and \S\ref{sec:lpt_pk} to compute the matter-shape power spectrum. This correlation contributes to the cosmic shear auto-spectrum as the correlation between intrinsic alignments and lensing (`GI'), and is also a subset of the contributions in galaxy-galaxy lensing through the galaxy-intrinsic alignment (`gI') cross correlation. We start with matter-shape correlations in the Zel'dovich approximation before moving on to the full 1-loop calculation. This section begins with a pedagogical derivation of the Lagrangian linear alignment contribution to the `GI' correlator in order to establish the notation and lines of thought used to evaluate the panoply of terms which contribute to the full one-loop expansion. 

\subsection{Matter-Shape Correlators in the Zel'dovich Approximation}
\label{subsec:za_gm}
As a warm-up to the general structure of the calculations we will carry out in the rest of this paper, we begin by examining only the terms which depend on the Zel'dovich displacement and the linear density and tidal fields. As seen in equation~\ref{eqn:lpt_spectra} we need to compute correlators of these fields ($\Psi^{(1)}, \delta, s_{ab}$) in Lagrangian coordinates $\bq$. These correlators between Lagrangian fields and displacements are the building blocks of predictions in LPT and have been well-studied and computed in previous works \cite{Carlson13}. Their definitions are reviewed in Appendix~\ref{app:lagcorrs}.

The linear matter-shape correlator in LPT is given by
\begin{equation}
    P_{ab} (\bk) = c_s \int d^3 \bq e^{i \bk \cdot \bq} \langle e^{i \bk \cdot \Delta(\bq_1, \bq_2)}\  s_{ab} (\bq_2) \rangle.
\end{equation}
Assuming the Zel'dovich approximation ($\bPsi = \bPsi^{(1)}$), from the generating function formalism we see that this is equivalent to
\begin{equation}
    P_{ab} (\bk) \ni c_s \int d^3 \bq e^{i \bk \cdot \bq - \frac12 k_i k_j A_{ij}(\bq)}\ i k_i B_{iab}(\bq).
\end{equation}
where we have followed ref.~\cite{Kokron22} and defined $B_{iab} (\bq) = \langle \Delta^{(1)}_i (\bq) s_{ab}(\bq_2) \rangle$. The angular structure of $B_{iab}$ was computed in ref.~\cite{White_2014} to be
\begin{equation}
    B_{iab}(\bq) = \hq_i \delta_{ab} \mathcal{J}_2 (q) + \big [ \delta_{ia}\hq_b + \delta_{ib}\hq_a \big]\mathcal{J}_3 (q) + \hq_i \hq_a \hq_b \mathcal{J}_4 (q)
\end{equation}
where $\mathcal{J}_n$ are sums of Hankel transforms of the linear power spectrum. From the discussion surrounding Equation~\ref{eqn:Qiab} it is clear that $B_{iab}$ can also be written in terms of the Legendre basis where it is manifest that only two of $\mathcal{J}_{2,3,4}$ are independendent due to the trace condition. As discussed in \S\ref{sec:lpt_pk} the resulting signal will have a quadrupole structure by symmetry, and projecting out this component by dotting it with $i k_i P_2^{ab}(\hk)$ gives us
\begin{equation}
\label{eqn:linearzapred}
     ik \int d^3 \bq\ e^{i \bk \cdot \bq - \frac12 k_i k_j A_{ij}(\bq)}  \left( -\frac{4}{15} \mu\ \xi^1_{-1} (q) + \frac{2}{5} P_3 (\mu) \xi^3_{-1} (q) \right),
\end{equation}
where we have defined $\mu = \hk \cdot \hq$, $P_\ell(\mu)$ is the $\ell$-th scalar Legendre polynomial and $\xi^\ell_n$ are the generalized correlation functions defined in \S\ref{ssec:notation}. In the absence of IR-resummation ($A_{ij} \rightarrow 0$) the angular integrals in the integral above can be done analytically as in Equation~\ref{eqn:tensorlegendrepair} and we simply obtain the linear theory prediction of $\frac23 P_{\rm lin}(k)$. \par

Let us now look at the contribution of quadratic operators in the Zel'dovich approximation. These come from the operators $\delta s_{ab}$ and $s_{ac} s_{cb}$. Noting that both of these operators are products of Gaussian fields we can directly apply Wick's theorem in configuration space to compute the relevant Lagrangian correlators, i.e. 
\begin{align}
\label{eqn:zadeltas}
    &-\frac{1}{2} k_i k_j \langle \Delta_i \Delta_j \delta (\bq_2) s_{ab}(\bq_2) \rangle_c = -k_i k_j \underbrace{U_i}_{\langle \Delta_i \delta\rangle} \underbrace{B_{jab}}_{\langle \Delta_j s_{ab} \rangle} \nonumber \\
     &-\frac{1}{2} k_i k_j \langle \Delta_i \Delta_j  (s^2_{ab})(\bq_2)\rangle_c = - k_i k_j  \underbrace{B_{iac}}_{\langle \Delta_i s_{ac} \rangle}  \underbrace{B_{jcb}}_{\langle \Delta_j s_{cb} \rangle}.
\end{align}
For the latter we also need to remove the trace of the right hand side, but this is accomplished automatically by projecting into the quadrupole $P_2^{ab}(\hk)$. Much like the linear contribution these terms can be expressed as products of $\xi^\ell_n$ and $\mu$ in the integrand.

Finally, let us examine the contributions of cubic bias operators built from $\{ \delta, s_{ab}\}$. As described in \S\ref{ssec:cubic} these operators' contributions are degenerate with the linear shape bias---let us see how this occurs in practice in the case of $\delta^2 s_{ab}$. In this case the only contribution at one loop is
\begin{equation}
    i k_i \avg{ \Delta_i \delta^2 s_{ab}(\textbf{0}) } = i k_i \avg{\Psi^{(1)}_i(\bq) (\delta^2 s_{ab})(\textbf{0})} = i \sigma^2 k_i \avg{\Psi^{(1)}_i(\bq) s_{ab}(\textbf{0})}
\end{equation}
which is equivalent to the linear contribution with $c_s = \sigma^2$ or, equivalently, can be removed by using the normal-ordered operator as we will do in this paper. The case of the other Zel'dovich cubic operators is entirely similar.

\subsection{Terms with One Pairwise Displacement ($\Delta$)}
\label{subsec:1delta}
In the previous section we computed contributions to the matter--shape power spectrum which arose solely due to the linear Zel'dovich displacement. At one loop in perturbation theory we will also find contributions from the second and third order displacements to these spectra. The contributions with one pairwise displacement $\Delta$ will be structured as
\begin{equation*}
    \langle O^{(4-n)}_{ab}(\bq_2)\ \Delta^{(n)}_i  \rangle = - \langle  O^{(4-n)}_{ab}(\bq_2)\ \Psi^{(n)}_i(\bq_1)   \rangle. 
\end{equation*}
This is precisely the correlator whose structure we derived using symmetry arguments in Equation~\ref{eqn:Qiab}. The contribution to the power spectrum is then computed from contraction of this basis with the hanging factor of $ik_i$ and the $P^2_{ab}(\hk)$ tensor, as previously carried out for the linear Zel'dovich correlator. 

Let us first consider terms due to the second-order displacement. For these (22) terms we can use that $\Psi^{(2)}_i$ is related to the inverse gradients of $\delta^2$ and $s^2$ such that
\begin{equation}
    \langle O^{(2)}_{ab}(\bk) | \Psi^{(2)}_i(\bk') \rangle' = -\frac{1}{7} \frac{i k_i}{k^2}  \langle O^{(2)}_{ab}(\bk) | (\delta^2 - \frac{3}{2} s^2)(\bk') \rangle' = -\frac{i k_i}{k^2} \Big( \frac{k_a k_b}{k^2} - \frac{1}{3} \delta_{ab} \Big) A(k)
    \label{eqn:O2Psi2}
\end{equation}
where the last step follows from the tracelessness of $O_{ab}$. From the above we see that $F_3 = -\frac{2i}{5k} A$ and $F_1 = \frac{2i}{15k} A$ are both described by a single scalar function. Since two of the quadratic spin operators are simply given by products of $\delta$ and $s_{ij}$ we may compute their correlators with the left-hand quantity by Wick contraction in configuration space. The resulting decompositions into generalized correlation functions are
\begin{equation}
\label{eqn:s2psi2}
    A_{s^2}(k) = - 4\pi \int dq\ q^2\ j_2(kq)\ \Big(\frac{4}{5145}\Big)\ \Big[ 49 \xi^0_0 \xi^2_0 + 130 (\xi^2_0)^2 + 36 \xi^2_0 \xi^4_0 - 45 (\xi^4_0)^2 \Big],
\end{equation}
\begin{equation}
\label{eqn:dspsi2}
    A_{\delta s} = - 4\pi \int dq\ q^2\ j_2(kq)\ \Big(-\frac{4}{245}\Big)\ \xi^2_0 \Big[ 14 \xi^0_0 + 5 \xi^2_0 - 9 \xi^4_0 \Big]. 
\end{equation}
The third, $L^{(2)}_{ab}$, is \textit{equal} to the left-hand side up to a few factors of $k$, so that the relevant spectrum is equivalent to the autospectrum of $\Psi^{(2)}$ which was already computed for the matter power spectrum in ref.~\cite{Matsubara08a}\footnote{In fact following the procedure above decomposing $\Psi^{(2)}$ and $L^{(2)}$ into a derivatives of a scalar we recover the FFTLog expression for $Q_1$ derived in \cite{Schmittfull_2016}.}:
\begin{equation}
\label{eqn:L2psi2}
    A_{L^{(2)}}(k) = -\frac{9}{98}  Q_1(k).
\end{equation}

Finally, let us examine the (13) terms. The contributions due to the cubic bias operators are equivalent to swapping in $A_{O}(k)$ for $P(k)$ when computing $B_{iab}$ for the Zel'dovich $(1, s_{ab})$ spectrum, as previously discussed. Similarly, contributions due to the third-order displacement can be computed by swapping
\begin{equation}
 \label{eqn:spsi3}
    A_s(k) = \frac{5}{21} R_1(k).
\end{equation}
with $P(k)$ when computing e.g. $\avg{\Delta_i^{(3)} s_{ab}}$ rather than $B_{iab}$.

\subsection{Terms with Two $\Delta$'s}
Certain spectra will receive contributions from two powers of the displacement $\Delta$. These spectra take the form
\begin{equation}
    \langle \Delta_i \Delta_j  O_{2,ab} \rangle = \langle  \Psi_{2,i} \Psi_{2,j} O_{2,ab}\rangle + \langle \Psi_{1,i} \Psi_{1,j} O_{2,ab} \rangle - \langle  \Psi_{1,i} \Psi_{2,j}O_{2,ab} \rangle - \langle \Psi_{2,i} \Psi_{1,j} O_{2,ab}  \rangle.
    \label{eqn:two_displacements}
\end{equation}
We see that there are three kinds of terms: the first one is a zero lag term, the second is a correlator of two displacements at one point correlated with a bias operator at a second, the third is the product of a displacement and bias operator correlated with another displacement. All three terms become identical in the zero-separation limit such that the total correlator vanishes.

As usual it will be easier to compute the Fourier transforms of these objects. Using the symmetry of the correlator between $i,j$ and $a,b$---note that in this case we only want the symmetrized component in $i,j$ since we will dot these with $k_i k_j$---we can write the Fourier transform in terms of the four-indexed orthogonal tensor components (\S\ref{ssec:angular})
\begin{equation}
    \langle  \Delta_i \Delta_j O_{2,ab} \rangle(\bk) = A_4(k) \mathcal{Q}^4_{ijab}(\hk) + A_2(k) \mathcal{Q}^{2A}_{ijab}(\hk)  + B_2(k) \mathcal{Q}^{2B}_{ijab}(\hk)  + A_0(k) \mathcal{Q}_{ijab}^0(\hk) 
\end{equation}
where we have defined
\begin{align}
    &\mathcal{Q}^4_{ijab}(\hk)  = P^{ijab}_4(\hk) \nonumber,\\
    &\mathcal{Q}^{2A}_{ijab}(\hk)  = P^{ij}_2(\hk) \delta_{ab} - \frac{3}{4} \Big(P^{ia}_2(\hk) \delta_{jb} + P^{ib}_2(\hk) \delta_{aj}+ P^{aj}_2(\hk) \delta_{ib} + P^{bj}_2(\hk) \delta_{ai}  \Big) \nonumber, \\
    &\mathcal{Q}^{2B}_{ijab}(\hk)  =  \delta_{ij} P^{ab}_2(\hk) \nonumber, \\
    &\mathcal{Q}^0_{ijab}(\hk)  = \delta_{ij} \delta_{ab} - \frac{3}{2} \Big( \delta_{ia} \delta_{jb} + \delta_{ja} \delta_{ib} \Big).
    \label{eqn:Q4basis}
\end{align}
The $\mathcal{Q}^\ell$ are orthogonal except for
\begin{equation*}
    \mathcal{Q}^{2A}_{ijab} \mathcal{Q}^{2B}_{ijab} = -\frac{9}{2}
\end{equation*}
so it is straightforward to obtain the various components by direct contraction. Note that it is important to subtract the zero-lag term from the $\mathcal{Q}^0$ piece after Fourier transforming to enforce the $\bq \rightarrow 0$ limit. We will now derive expressions for the different two-displacement contributions for linear and quadratic alignment operators.

\subsubsection{Contribution to linear alignment -- $\langle
s_{1, ab} \Delta^{(1)}_i \Delta^{(2)}_j\rangle$}
For the linear alignment operator we will have contributions from the second-order displacement vector. The different terms in Equation~\ref{eqn:two_displacements} are given by
\begin{equation*}
    \langle s_{ab} (\bk) \; | \; (\Psi^{(1)}_i \Psi^{(2)}_j)(\bk') \rangle' = 2 S_{ab}(\bk) P(k) \int_\bp L^{(2)}_i(\bp,-\bk) L^{(1)}_j(-\bp)\  P(p)
\end{equation*}
\begin{equation*}
    \langle (\Psi^{(1)}_i s_{ab})(\bk)\; | \Psi^{(2)}_j(\bk') \rangle' = 2  \int_\bp L^{(1)}_i(\bp) S_{ab}(\bk-\bp) L^{(2)}_j(-\bp,-\bk+\bp) P(p) P(|\bk-\bp|).
\end{equation*}
\begin{equation}
\label{eqn:psi2s|psi1}
    \langle (\Psi^{(2)}_i s_{ab})(\bk)\; | \Psi^{(1)}_j(\bk') \rangle' = 2 L^{(1)}_j(-\bk) P(k) \int_\bp L^{(2)}_i(\bp,\bk)\ S_{ab}(-\bp) P(p)
\end{equation}
where we have defined $S_{ab}(\bk) = (\hk_a \hk_b - \frac13 \delta_{ab}$). In order to maintain the symmetry properties of the basis above we need to evaluate the symmetrized version $\langle
s_{1, ab} \Delta^{(1)}_{(i} \Delta^{(2)}_{j)}\rangle$, which is in fact what enters into the power spectrum since each $\Delta$ is always contracted with an identical $k$. Each term is computed by a separate technique that we have already introduced -- reduction to a previous result, reduction to configuration-space Wick contractions or direct evaluation into sums of products of generalized correlation functions. 

The first term -- $\langle s_{ab} (\bk) \; | \; (\Psi^{(1)}_i \Psi^{(2)}_j)(\bk') \rangle'$-- can be inferred from a related correlator in the galaxy power spectrum \cite{Matsubara08b}
\begin{align}
\label{Mastsubara}
    \langle s_{ab} (\bk) \; | \; (\Psi^{(1)}_i \Psi^{(2)}_j)(\bk') \rangle' &= \Big( \frac{k_a k_b}{k^2} - \frac{1}{3} \delta_{ab} \Big) \langle \delta(\bk) | (\Psi^{(1)}_i \Psi^{(2)}_j)(\bk') \rangle' \nonumber \\
    &= - \Big( \frac{k_a k_b}{k^2} - \frac{1}{3} \delta_{ab} \Big) \Big( -\frac{3}{14} \frac{\delta_{ij}}{k^2} R_1(k) + \frac{3}{14} \frac{k_i k_j}{k^4} (R_1 + 2 R_2) \Big).
\end{align}
The second two terms both involve correlators of products of displacements and bias operators with a single displacement; since displacements in LPT are inverse gradients we can pull out a power of $i k^{-1} \hk$ from both of them such that
\begin{align}
    \langle (\Psi^{(1)}_i s_{ab})(\bk)\; | \Psi^{(2)}_j(\bk') \rangle' &= \frac{k_j}{k^2} \Big( Q^s_3(k) \mathcal{Q}^3_{iab}(\hk) - Q^s_1(k)\mathcal{Q}^1_{iab}(\hk) \Big) \nonumber \\
    \langle (\Psi^{(2)}_i s_{ab})(\bk)\; | \Psi^{(1)}_j(\bk') \rangle' &=  \frac{\,k_j}{k^2}  \left [ R_3^{s}(k) \mathcal{Q}^3_{iab} + R_1^{s}(k) \mathcal{Q}^1_{iab} \right ].
    \label{eqn:ODD_13form} 
\end{align}
The first of these can be evaluated using the inverse-gradient form of $\bPsi^{(2)}_i$ as in Equation~\ref{eqn:O2Psi2}. Explicit expressions for these kernels can be found in Appendix~\ref{app:integral_details}. It is worth noting that these two integrals are not independent, in that $R_3^s = \frac{1}{4} A_{st} - \frac{1}{2} R_1^s$. Recasting into the four-indexed $\mathcal{Q}_{ijab}$ basis we find the final expression for the components of the two-displacement linear alignment correlator.
\begin{align*}
    &A_0^s(k) = \frac{2}{105 k^2} (R_1 + 2R_2) + \frac{2}{3k} (Q^s_1 - R^s_1) \\
    &A_2^s(k) = \frac{8}{147 k^2} (R_1 + 2R_2) + \frac{4}{21k} (7 Q^s_1 + Q^s_3 - 7R^s_1 + R^s_3) \\
    &B_2^s(k) = \frac{4}{49k^2} (3R_1 - R_2) - \frac{10}{21k} (Q^s_3 + R^s_3) \\
    &A_4^s(k) = -\frac{24}{245k^2} (R_1 + 2R_2) - \frac{8}{7k} (Q^s_3 + R^s_3).
\end{align*}
\begin{figure}
    \centering
    \includegraphics[width=\textwidth]{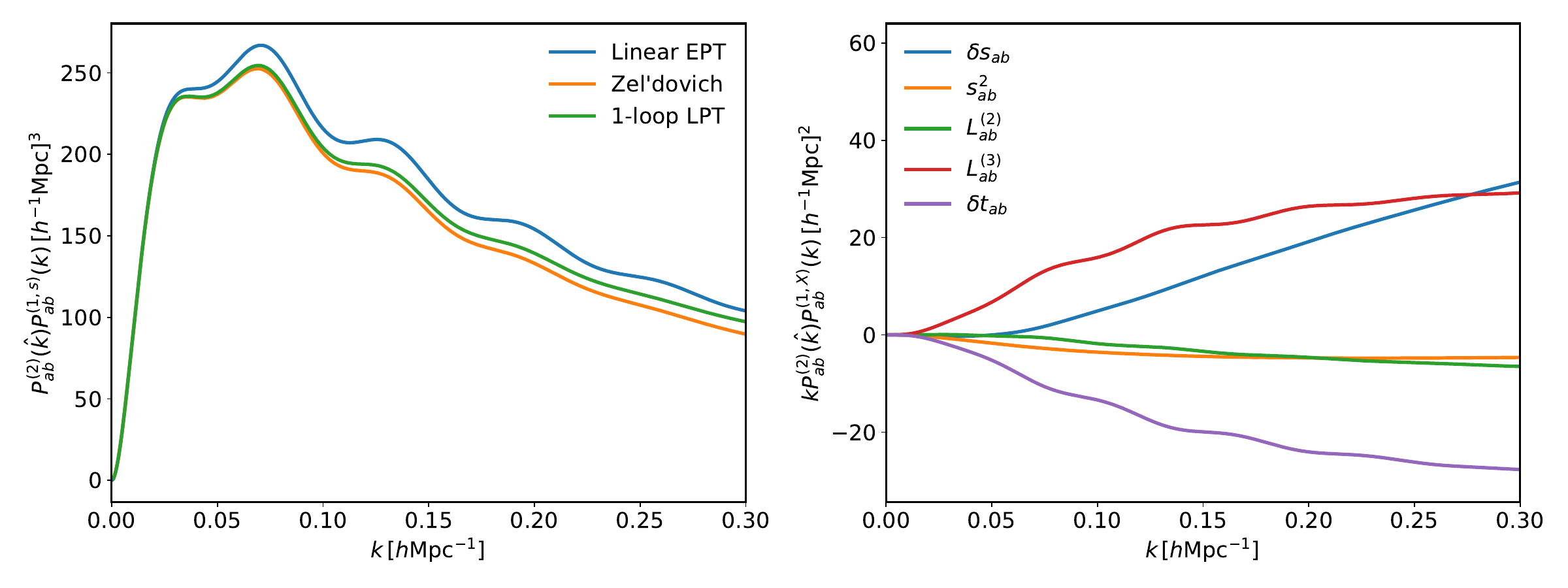}
    \caption{Contributions to the matter--shape spectrum at one-loop order. \emph{Left:} The individual contributions to the matter-tide correlator compared to the linear Eulerian prediction (in blue), the Zel'dovich approximation (orange) and the 1-loop LPT spectrum including higher-order displacements. The corrections are $\sim 20\%$ at $k=0.3 \ihmpc$. \emph{Right:} The remaining matter--shape spectra at one-loop order, with all shape bias coefficients set to one.}
    \label{fig:densshapepk}
\end{figure}

\subsubsection{Quadratic Bias}

We now proceed to compute the correlations between quadratic alignment operators and two $\Delta$ operators. There are three operators at quadratic order
\begin{equation*} 
O^{(2)}_{ab} \ni (s_{ak}s_{kb}, \delta s_{ab}, L^{(2)}_{ab}).
\end{equation*}
The first two of these are straightforward products of linear operators and their contributions were already obtained in Equation~\ref{eqn:zadeltas}. The third one, $L^{(2)}_{ab}$, however is nonlocal and we have to compute the (13) and (22) integrals
\begin{align}
    \langle (\Psi^{(1)}_i L^{(2)}_{ab})(\bk)\; | \Psi^{(1)}_j(\bk') \rangle' &= 2 L^{(1)}_i(-\bk) P(k) \int_\bp L^{(1)}_j(\bp)\ L^{(2)}_{ab}(-\bp,\bk) P(p) \nonumber \\
    \langle L^{(2)}_{ab}(\bk) \; | \;  (\Psi^{(1)}_i \Psi^{(1)}_j)(\bk') \rangle' &= 2 \int_\bp L^{(1)}_i(\bp) L^{(1)}_j(\bk-\bp)\ L^{(2)}_{ab}(\bp,\bk-\bp) P(p) P(|\bk-\bp|)     
    \label{eqn:L2Psi1xPsi1}
\end{align}
where we have dropped a term using $L^{(2)}_{ab}(\bp,-\bp)=0$. We can decompose the former into the form of Equation~\ref{eqn:ODD_13form} where in this case we substitute in the corresponding kernels $R_{1,3}^{L^{(2)}}$ whose explicit definitions are given in Appendix~\ref{app:integral_details}. Again, these integrals are not independent, but are rather linear combinations of previous cubic contributions, $k R_1^{L^{(2)}} = -1/10 A_{\delta t} + \frac{3}{5} A_{st}$ and $k R_1^{L^{(2)}} = -1/5 A_{\delta t} - \frac{3}{10} A_{st}$.

For the latter, we note these are equivalent to results obtained when computing cubic correlators of displacements \cite{Matsubara08b}
\begin{equation}
    \langle L^{(2)}_{ab}(\bk)  \; | \; (\Psi^{(1)}_i \Psi^{(1)}_j)(\bk') \rangle' = \frac{3}{14} \Big(\frac{-k_a k_b}{k^4}\Big) \Big( Q_1(k) \delta_{ij} - (Q_1(k) + 2 Q_2(k)) \hk_i \hk_j \Big).
\end{equation}
Both of these integrals can then be Fourier-transformed back into real space by first projecting into the $\mathcal{Q}$ basis in Equation~\ref{eqn:Q4basis} as done for the linear alignment contribution. Combining again all the terms we have
\begin{align*}
    &A_0^L(k) = -\frac{1}{105 k^2} (Q_1 + 2Q_2) - \frac{2}{3k} R_1^{L^{(2)}} \\
    &A_2^L(k) = -\frac{4}{147 k^2} (Q_1 + 2Q_2) - \frac{4}{21k} (7 R^{L^{(2)}}_1 - R^{L^{(2)}}_3 ) \\
    &B_2^L(k) = -\frac{2}{49k^2} (3Q_1 - Q_2) - \frac{10}{21k} R^{L^{(2)}}_3 \\
    &A_4^L(k) = +\frac{12}{245k^2} (Q_1 + 2 Q_2) - \frac{8}{7k} R^{L^{(2)}}_3
\end{align*}
This concludes the derivation of the matter--shape spectrum to one-loop order in perturbation theory. In Figure~\ref{fig:densshapepk} we show predictions for all matter--shape spectra derived above at $z = 0.51$, chosen to match the N-body measurements in \S\ref{sec:nbody}, as computed by \texttt{spinosaurus}, the code we present as part of this publication.

\subsection{IR Resummation and Shape Correlation Functions}

\begin{figure}
    \centering
    \includegraphics[width=\textwidth]{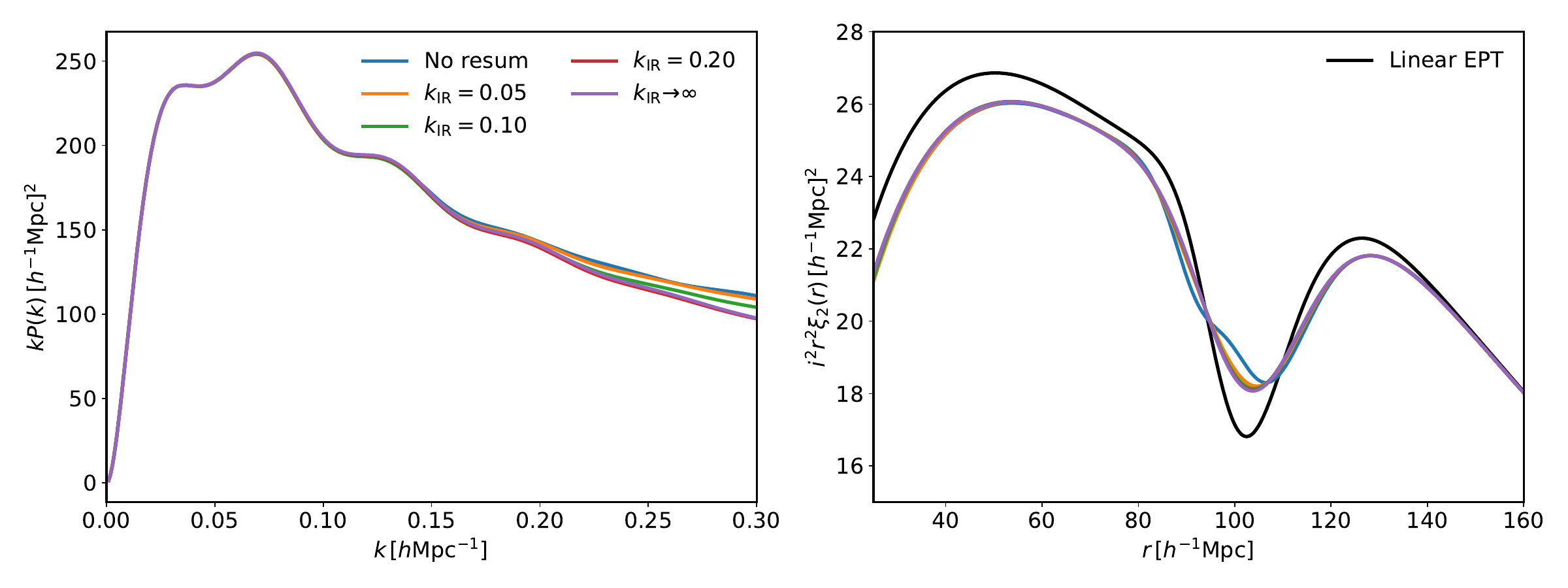}
    \caption{The impact of the resummation scale $k_{\rm IR}$ on density--shear observables from 1-loop (non-resummed) EPT to fully-resummed LPT. \emph{Left:} the matter--shape power spectrum as $k_{\rm IR}$ is changed, assuming linear Lagrangian shape bias. \emph{Right:} The matter--shape correlation function quadrupole as a function of resummation scale, compared to the linear alignment prediction. Note the failure of the un-resummed theory in capturing the BAO feature, as well as the insensitivity of this feature to $k_{\rm IR}$ in LPT.}
    \label{fig:baodip}
\end{figure}
As previously mentioned, a key advantage of formulating a theory of intrinsic alignments within the framework of LPT is the ability to naturally capture the smoothing of the BAO feature due to long-wavelength displacements. To illustrate this effect on intrinsic alignment observables, we calculate the matter--linear shape power spectrum and associated correlation function (given by $\xi^2_0$) in comparison to linear theory and the un-resummed 1-loop prediction, shown in Figure~\ref{fig:baodip}. The BAO features manifests as a ``dip'' in the correlation function since the signal has a quadrupolar angular dependence, much like the BAO feature in the redshift-space quadrupole of the galaxy density. We show these spectra as the IR resummation scale, $k_{\rm IR}$, is varied from no-resummation to fully resummed LPT. While the broadband shape of the power spectrum is mildly affected by $k_{\rm IR}$ at small scales, a fit to data in practice would adjust for EFT corrections (stochastic and counterterms) that would make up for some of the broadband difference. The impact of this resummation on the BAO feature and correlation function is on the other hand stark, with the BAO damped in the resummed versions but persisting out to high $k$ and leading to wiggly features in configuration space when unresummed. The damping is realtively robust to the choice of resummation scale, which plays little effect on the configuration-space BAO feature shape, reflecting that most of the damping is done by long-wavelength modes near the BAO scale.

\section{Galaxy Density-Shape Cross Spectrum}
The final scalar--shape set of correlators we must consider are the galaxy density--shape cross-spectra. These spectra are the building blocks for the `gI' contribution to intrinsic alignments in studies of galaxy--galaxy lensing, as well as studies cross-correlating 3D catalogs of galaxy shapes with their positions. These cross-correlations will be of the form $(\delta_g,  M_{ab}).$
In the `linear alignment' paradigm the only operators that contribute to this cross-correlation are linear in density -- $\delta$ and $s_{ab}$. However, at 1-loop this cross-spectrum relies also on correlations of higher order operators, whose cross correlations with density bias operators must be included in a complete model of shape statistics. We proceed to enumerate these below. Since each bias operator includes at least one additional power of the density field, the calculations are simpler and the vast majority can be done via configuration-space Wick contractions. 
\label{sec:densshape}

\subsection{Terms with no displacements}
Let us list contributing spectra in terms of ordered pairs $O_a(\bq_2),O_b(\bq_1)$, giving expressions of the cross-spectra for each pair. The linear term is simply
\begin{itemize}
    \item $(\delta, s_{ab})$: $E_{ab} = -\frac{2}{3} P^2_{ab}(\hq) \xi^2_0.$
\end{itemize}
These bias operators also produce a term due to the square of the first cumulant (Eqn.~\ref{eqn:genfunc}) $\langle \delta(\bq_1) s_{ab}(\bq_2) \Delta_i \Delta_j \rangle = 2 U_i  B_{jab}$. This is the only contribution due to two pairwise displacements in the non-matter density-shape correlators.

For pure contributions due to two quadratic bias operators we have the following
\begin{itemize}
    \item $(\delta s_{ab}, \delta^2): \ 2 \xi E_{ab}$
    \item $(s^2_{ab}, \delta^2): \ 2E_{ac} E_{cb}$
    \item $(\delta s_{ab}, s^2): \ 2 E_{cd} C_{abcd}$
    \item $(s^2_{ab}, s^2): \ 2 C_{acde} C_{cbde}$.    
\end{itemize}
There are in addition to the above two contributions from $L^{(2)}_{ab}$ that can be computed as Hankel transforms of LPT kernels using that $L_{ab} = \partial_i \Psi_j$ \cite{Vlah_2016}:
\begin{itemize}
    \item $(L^{(2)}_{ab},\delta^2): \ \frac{2}{7} P_2(\hq)\xi^2_0 \left[ Q_8 \right]$
    \item $(L^{(2)}_{ab},s^2): \ \frac{2}{21} P_2(\hq) \xi^2_0 \left[ Q_{s^2} \right]$
\end{itemize}
where we have dropped the trace components. Finally, the cross correlation of cubic bias operators with the linear shear field is given by computing $E_{ab}$ using $A_O(k)$.
\begin{figure}
    \centering
    \includegraphics[width=\textwidth]{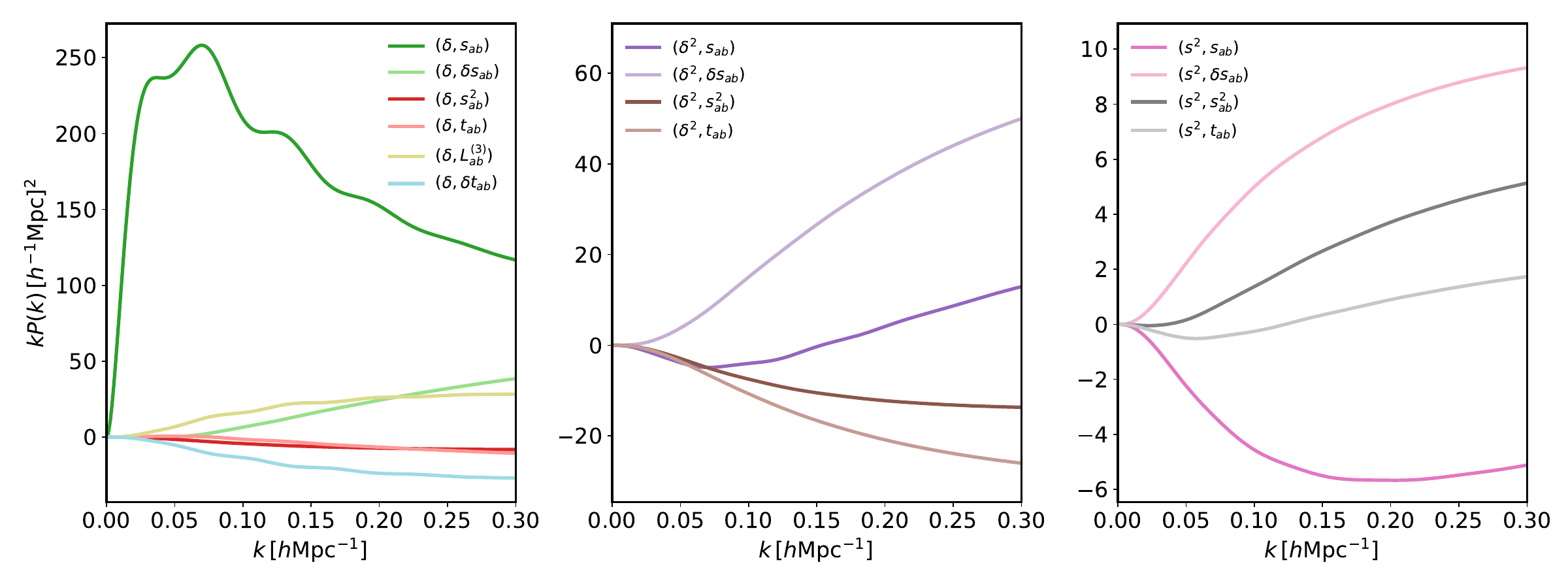}
    \caption{The full set of density--shape correlators at one-loop order. Note there is significant variation in y-axes between panels. \emph{Left:} Spectra between the linear density operator and galaxy shape operators. \emph{Center:} Spectra between the quadratic density operator and shape operators.  \emph{Right:} Spectra between the tidal field strength and shape operators. }
    \label{fig:dens_shape}
\end{figure}
\subsection{Terms with One $\Delta$}
In the case of one pairwise linear displacement we have the following contributions
\begin{itemize}
    \item $(s_{ab}, \delta^2): 2 i k_i U_i E_{ab} $
    \item $(s_{ab}, s^2): 2 i k_i B_{icd} C_{abcd} $
    \item $(\delta s_{ab}, \delta): i k_i (\xi B_{iab} + U_i E_{ab}) $
    \item $(s^2_{ab}, \delta): 2 i k_i B_{iac} E_{cb}$
    \item $(L^{(2)}_{ab}, \delta): \langle L^{(2)}_{ab,2} \Psi^{(1)}_{i,2} \delta_1 \rangle -  \langle L^{(2)}_{ab,2} \Psi^{(1)}_{i,1} \delta_1 \rangle$.
    \item $(s_{ab}, \delta): \langle s_{ab}(\bq_2) \Psi_{i}^{(2)}(\bq_2) \delta (\bq_1) \rangle - \langle s_{ab}(\bq_2)\Psi_{i}^{(2)}(\bq_1) \delta(\bq_1) \rangle $
\end{itemize}
Both terms in the penultimate expression involve previously-computed integrals. In particular using Equation~\ref{eqn:L2Psi1xPsi1} we have
\begin{equation*}
     \langle L^{(2)}_{ab,2} \Psi^{(1)}_{i,2} \delta_1 \rangle  =  \xi^3_0\left[R_3^{L^{(2)}}(k)\right] \mathcal{Q}^3_{iab}(\hq) - \xi^1_0\left[R_1^{L^{(2)}}\right] \mathcal{Q}^1_{iab}(\hq)
\end{equation*}
and from correlators previously computed for galaxy densities \cite{Matsubara08b} we have
\begin{equation}
    \langle L^{(2)}_{ab,2} \Psi^{(1)}_{i,1} \delta_1 \rangle = \frac{3}{7} \Big( \frac{2}{5} \xi^3_{-1}\left[Q_5\right] \mathcal{Q}^3_{iab}(\hq) + \frac{2}{15} \xi^1_{-1}\left[Q_5\right] \mathcal{Q}^1_{iab}(\hq) \Big).
\end{equation}
The final item is a contribution due to the second-order displacement
given by
\begin{align}
    \langle s_{ab,2} \Psi_{i,2}^{(2)} \delta_1 \rangle &= \xi^3_0\left[R_3^{s}\right] \mathcal{Q}^3_{iab}(\hq) - \xi^1_0\left[R_1^{s}\right] \mathcal{Q}^1_{iab}(\hq) \nonumber \\
    \langle s_{ab,2} \Psi_{i,1}^{(2)} \delta_1 \rangle &= \frac{3}{7} \Big( -\frac{2}{5} \xi^3_{-1}\left[R_1+R_2\right] \mathcal{Q}^3_{iab}(\hq) - \frac{2}{15} \xi^1_{-1}\left[R_1+R_2\right] \mathcal{Q}^1_{iab}(\hq) \Big)
    \label{eqn:spsi2d}
\end{align}
where in the first line we have used our result from Equation~\ref{eqn:ODD_13form}. This concludes the derivation of all density--shape correlators which contribute at one-loop order. We show these spectra, computed at $z=0.51$, in Figure ~\ref{fig:dens_shape}. Except for the pure linear-Lagrangian bias term $(\delta,s_{ab})$ spectrum, all the other spectra are 1-loop at leading order and amount to perturbative corrections to the overall clustering signal over the range shown.%

\section{Shape--shape correlations} 
\label{sec:shapeshape}

Finally let us compute the shape-shape two-point function of galaxies. These correlations are particularly relevant for analyses of cosmic shear since they directly contribute to shear-shear correlations through the measured ellipticities of galaxies. Unlike in the density-shape correlations calculated in the previous two subsections these shape-shape correlations involve three distinct components given by the independent spectra of the $m=0,1,2$ helicity components. As was the case for the galaxy density-shape correlators all of the Lagrangian correlators needed for these calculations can be obtained by re-arranging previous results.  

\subsection{Advection Mixes Modes}
Before we enumerate all the contributions to shape-shape spectra at 1-loop let us first return to the illustrative example of linear Lagrangian shape bias in the Zel'dovich approximation, where all large-scale correlations are due to the Lagrangian shear operator $s_{ab}(\bq)$. This operator has only $m=0$ modes in Lagrangian space by definition.

\begin{figure}
    \centering
    \includegraphics[width=0.7\textwidth]{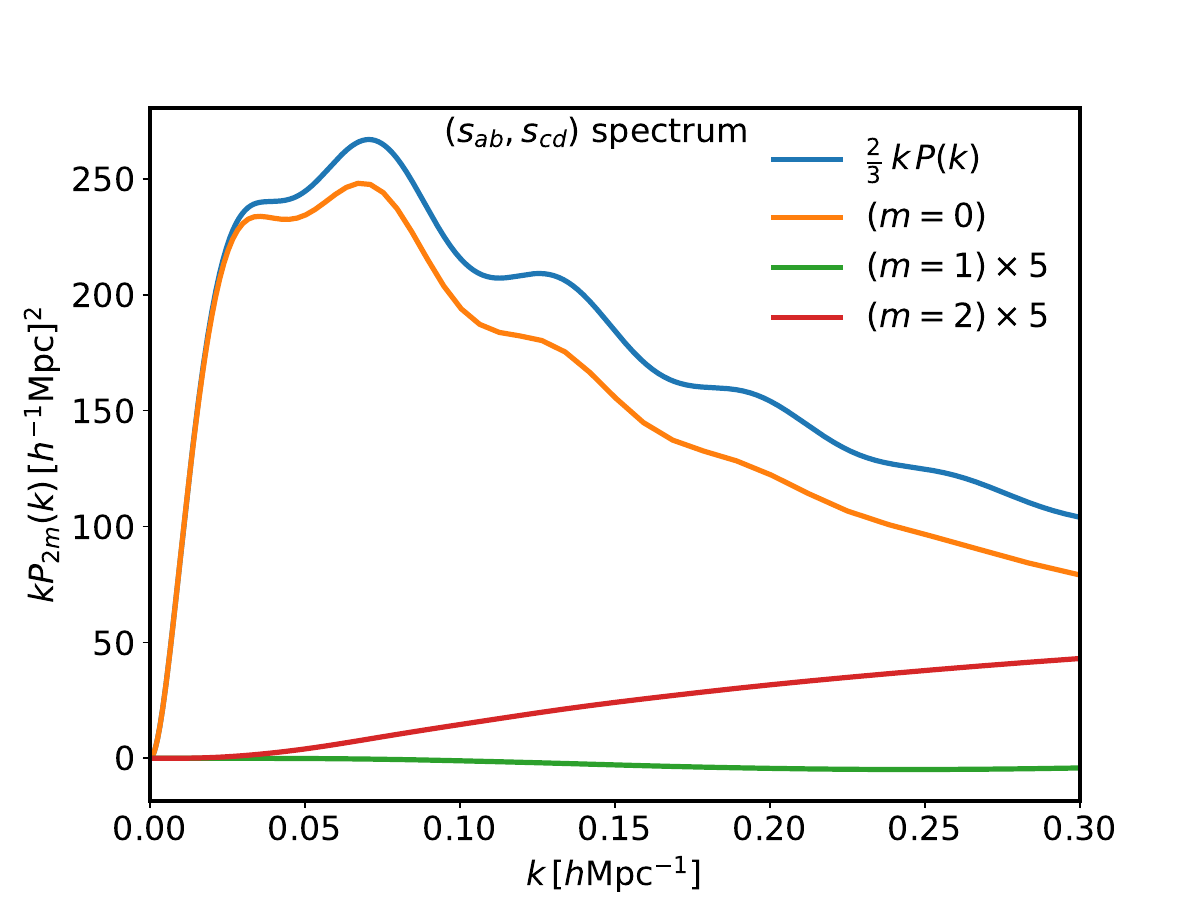}
    \caption{The linear shear--shear correlator pre and post-advection. The blue line shows the $m=0$ (and only) contribution to the pre-advection correlator and the subsequent lines show the generation of helicity 1 and 2 spectra from advection and non-linear evolution. The correlators are computed with $k_{\rm IR} = 0.2 \ihmpc$.}
    \label{fig:ss_advect}
\end{figure}

Let us examine the generation of nonlinearities due to advection. For simplicity we will operate in the Zel'dovich approximation and resum all linear displacements (i.e. $k_{\rm IR} = \infty$). In this regime the leading contribution is
\begin{equation}
    P_{s_{ab} s_{cd}}= \int d^3\bq\ e^{i\bk\cdot\bq - \frac{1}{2}k_ik_j A_{ij}} C_{abcd}(\bq).
\end{equation}
In order to extract the helicity components we need to project the tensor expression above into the helicity basis as in Equation~\ref{eqn:helicity_components}. This projection leads to angular integrals that can be written in terms of Legendre polynomials of $\mu = \hk \cdot \hq$
\begin{itemize}
    \item $m = 0$: $\int d^3\bq\ e^{i\bk\cdot\bq - \frac{1}{2}k_ik_j A_{ij}} \big( \frac{2}{15} \xi^0_0(q) - \frac{4}{21} \xi^2_0(q) P_2(\mu) + \frac{12}{35} \xi^4_0(q) P_4(\mu) \big)$
    \item $m = 1$:  $- \frac12 \int d^3\bq\ e^{i\bk\cdot\bq - \frac{1}{2}k_ik_j A_{ij}} \big( -\frac{4}{15} \xi^0_0(q) + \frac{4}{21} \xi^2_0(q) P_2(\mu) + \frac{16}{35} \xi^4_0(q) P_4(\mu) \big)$
    \item $m = 2$: $\frac12 \int d^3\bq\ e^{i\bk\cdot\bq - \frac{1}{2}k_ik_j A_{ij}} \big( \frac{4}{15} \xi^0_0(q) + \frac{8}{21} \xi^2_0(q) P_2(\mu) + \frac{4}{35} \xi^4_0(q) P_4(\mu) \big)$
\end{itemize}
In the absence of the displacements in $A_{ij}$, Equation~\ref{eqn:tensorlegendrepair} gives that the $m=1,2$ components would vanish, as they do in linear theory. With advection, however, they will tend to be nonzero, as shown in Figure~\ref{fig:ss_advect}. Indeed, the generation of nonzero helicity modes by advection represents a sizeable contribution at the tens-of-percents level compared to the linear theory prediction. This is analogous to the generation of B modes by CMB lensing \cite{Zaldarriaga98}. The contributions of cubic spin operators is again accomplished by simply exchanging $\xi^\ell_n \rightarrow \xi^\ell_n\left[A_O(k)\right]$ in the above.

The angular structure above is fully general to any advected correlator $f_{abcd}$ between two shape operators. Fundamentally this is because any such operators can be decomposed into the three Legendre basis tensors
\begin{align}
    &\mathcal{Q}^4_{abcd} = P_4^{abcd}(\hq) \nonumber\\
    &\mathcal{Q}^2_{abcd} = \delta_{ab} P_2^{cd}(\hq) + \delta_{cd} P_2^{ab}(\hq) - \frac{3}{4} \left( \delta_{ac} P_2^{bd}(\hq) + \delta_{ad} P_2^{bc}(\hq) + \delta_{bd} P_2^{ac}(\hq) + \delta_{bc} P_2^{ad}(\hq)\right) \nonumber\\
    &\mathcal{Q}^0_{abcd}(\hq)  = \delta_{ab} \delta_{cd} - \frac{3}{2} \Big( \delta_{ac} \delta_{bd} + \delta_{ad} \delta_{bc} \Big)
    \label{eqn:Qabcd}
\end{align}
as in \S\ref{ssec:angular}, which in this case is equivalent to noting that there are three independent helicity components describing any such correlator. These basis tensors are orthogonal to their Fourier-space counterparts $\mathcal{Q}^\ell_{abcd}(\hk)$ when $\ell \neq \ell'$ and dot into Legendre polynomials $P_\ell(\mu)$ otherwise, implying that the helicity spectra are each composed of the same three functions multiplying Legendre polynomials with coefficients depending only on $m$.

\subsection{$(s_{ab},s_{cd})$}
Beyond the lowest order contribution we have two further contributions
\begin{equation}
    P_{s_{ab} s_{cd}}= \int d^3\bq\ e^{i\bk\cdot\bq - \frac{1}{2}k_ik_j A_{ij}} \Big( i k_i \langle \Delta_i^{(2)} s_{2,ab} s_{1,cd} \rangle -  k_i k_j B_{iab} B_{jcd} \Big)
\end{equation}
The former term is given by
\begin{align*}
\langle \Delta_i^{(2)} s_{2,ab} s_{1,cd} \rangle &= \langle \Psi_i^{(2)}(\bq_2) s_{ab}(\bq_2) s_{cd}(\bq_1) \rangle - \langle \Psi_i^{(2)}(\bq_1) s_{ab}(\bq_2) s_{cd}(\bq_1) \rangle \\
&= \langle \Psi_i^{(2)}(\bq_2) s_{ab}(\bq_2) s_{cd}(\bq_1) \rangle + \langle \Psi_i^{(2)}(\bq_2) s_{cd}(\bq_2) s_{ab}(\bq_1) \rangle
\end{align*}
and can be expressed in terms of the integrals in Equation~\ref{eqn:ODD_13form}
\begin{align}
    \langle \Psi_i^{(2)} s_{ab} | s_{cd} \rangle' = i \left[ R^s_3(k) \mathcal{Q}^3_{iab}(\hk) + R^s_1(k) \mathcal{Q}^1_{iab}(\hk) \right] \left( \frac{k_c k_d}{k^2} - \frac{1}{3} \delta_{cd} \right).
\end{align}
In order to Fourier-transform this correlator, symmetrized between the two pairs of indices $(a,b)$, $(c,d)$, we can construct again a Legendre basis for rank-5 tensors satisfying that the last four indices are composed of two traceless and symmetric pairs using the method described in \S~\ref{ssec:angular}. This produces a set of five basis tensors $\mathcal{Q}^{5},\mathcal{Q}^{3A},\mathcal{Q}^{3B},\mathcal{Q}^{1A},\mathcal{Q}^{1B}$ whose construction follows the procedure described in \S~\ref{ssec:angular}\footnote{As this basis is composed of a large number of terms, we present its construction and manipulation in the \texttt{basis\textunderscore iabcd.nb} Mathematica notebook in the \texttt{spinosaurus} git repository.}. Note that this term has no helicity one or two contributions in EPT due to the factors of $\hk_a \hk_b - \delta_{ab}/3$, so that the only Eulerian contribution comes from the square of $B_{iab}$.
\begin{figure}
    \centering
    \includegraphics[width=\textwidth]{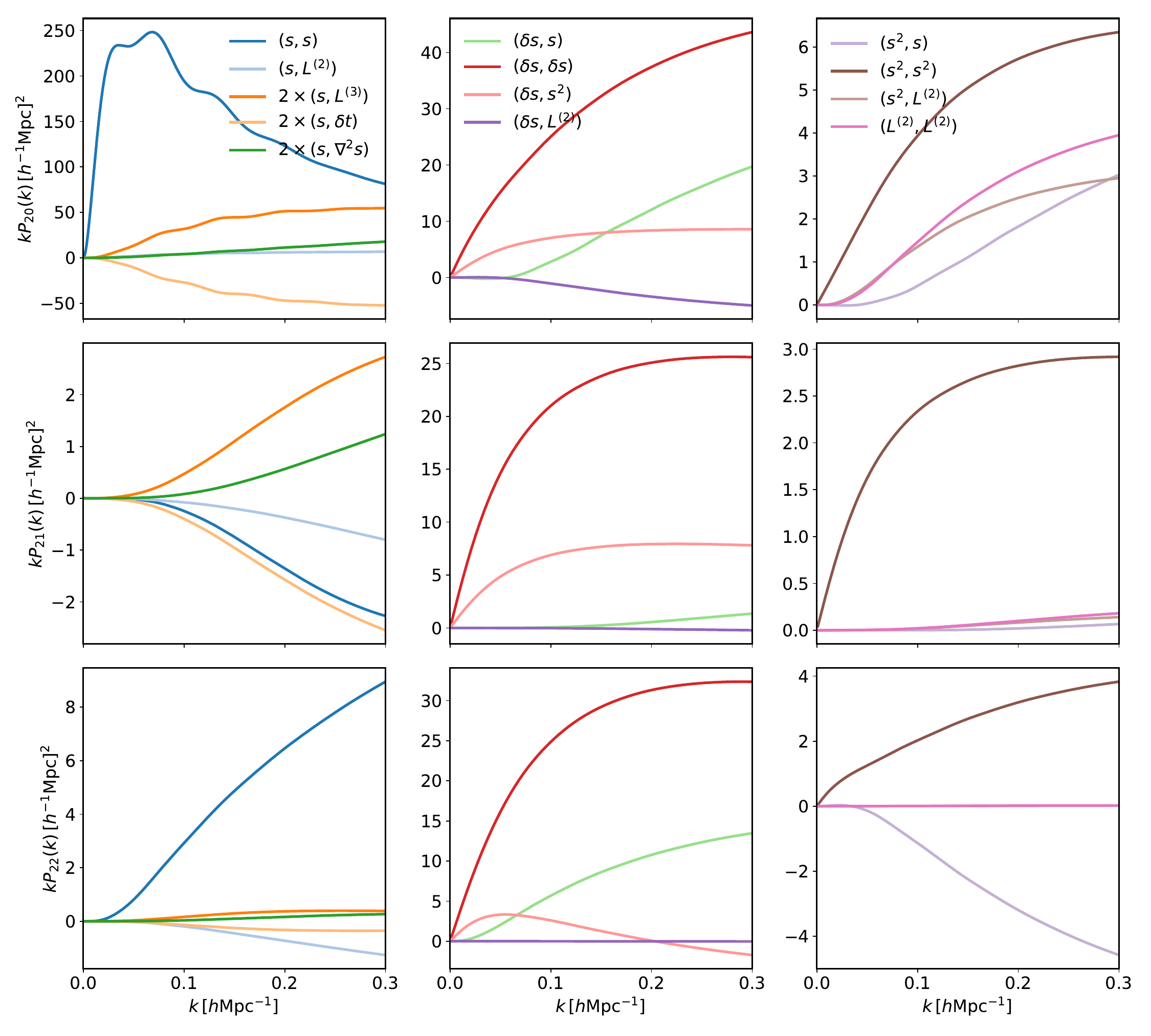}
    \caption{All components of shape--shape spectra at one-loop order. Each row shows a separate helicity $m=\{0, 1, 2\}$ of the shape--shape correlators, in descending order. }
    \label{fig:shapeshapeplot}
\end{figure}
\subsection{Linear x Quadratic}

Let us now look at correlators of the form
\begin{equation}
    i k_i \avg{\Delta_i s_{2, ab} O^{(2)}_{1, cd}} = i k_i \left(\avg{\Psi_{2,i} s_{2, ab} O^{(2)}_{1, cd}} - \avg{ s_{2, ab} \Psi_{1,i} O^{(2)}_{1, cd}} \right).
\end{equation}
Two of these are straightforward to write down in terms of Zel'dovich correlators
\begin{itemize}
    \item $(s_{ab}, \delta s_{cd})$: $i k_i (U_i C_{abcd} + E_{ab} B_{icd})$
    \item $(s_{ab}, s^2_{cd})$: $i k_i (C_{abce} B_{ied} + C_{abed} B_{iec})$
\end{itemize}
where we need also to remove the $cd$ trace of the latter. Removing the $c,d$ trace ensures that the resulting correlator is symmetric under exchange of $ab$, $cd$, though this is also accomplished directly when projecting into the helicity basis. Additionally, these operators do not have $m=1$ helicity modes prior to advection, as was first pointed out by ref.~\cite{Akitsu23}. Finally we have 
\begin{equation}
    (s_{ab}, L^{(2)}_{cd}):\ i k_i \left\langle s_{ab}(\bq_1) L^{(2)}_{cd}(\bq_2) \Delta_i \right\rangle = i k_i \left( \left\langle \Psi_i(\bq_2) L^{(2)}_{cd}(\bq_2) s_{ab}(\bq_1) \right\rangle + \left\langle \Psi_i(\bq_2) s_{ab}(\bq_2) L^{(2)}_{cd}(\bq_1) \right\rangle \right)
\end{equation}
Both of the correlators above can be related to ones we have already previously computed (Eqn.~\ref{eqn:L2Psi1xPsi1})
\begin{align*}
    &\avg{ (\Psi_i s_{ab})(\bk)  | \TF{L^{(2)}}_{cd}(\bk') }' = - i \left( Q^s_3(k) \mathcal{Q}^3_{iab}(\hk) - Q^s_1(k) \mathcal{Q}^1_{iab}(\hk)\right) \left( \frac{k_c k_d}{k^2} - \frac13 \delta_{cd} \right) \\
   &\avg{(\Psi_i \TF{L^{(2)}}_{cd})(\bk) |  s_{ab}(\bk') }' = i \left( R^{L^{(2)}}_3(k) \mathcal{Q}^3_{icd}(\hk) + R^{L^{(2)}}_1(k) \mathcal{Q}^1_{icd}(\hk)\right) \left( \frac{k_a k_b}{k^2} - \frac13 \delta_{ab} \right)
\end{align*}
where we have explicitly written the trace-free component in both cases. In order to project these onto the $Q^\ell_{iabcd}$ basis, it is necessary to further symmetrize the trace-removed correlators above under pair swap---while these correlators are not themselves symmetric under pair swap, their Fourier transform after dotting with $k_i$ and IR resummation must be due to parity\footnote{It is straightforward to check that the Fourier-space correlators above are also symmetric in this way after contracting with $k_i$.}, so this step can be safely performed, though it is also sufficient to project with the $Q^\ell_{iabcd}$ basis directly to remove offending components. Note also that these spectra again have no helicity one or two contributions in EPT due to the factors of $\hk_a \hk_b - \delta_{ab}/3$.

\subsection{Quadratic x Quadratic}

Finally let us consider the case of terms involving two quadratic bias operators. In this case at 1-loop order no further displacements can be involved so we need only consider Lagrangian correlators of the form $\avg{O_{1,ab} O_{2,cd}}$ whose angular structure is described by the basis in Equation~\ref{eqn:Qabcd}.

Several of the terms with two quadratic operators can be written purely in terms of Zel'dovich correlators
\begin{itemize}
    \item $(\delta s_{ab}, \delta s_{cd})$: $\xi C_{abcd} + E_{ab} E_{cd}$
    \item $(\delta s_{ab}, s^2_{cd})$: $E_{ce} C_{abed} + E_{ed} C_{abce}$
    \item $(s^2_{ab}, s^2_{cd})$: $C_{aecf} C_{ebfd} + C_{aefd} C_{ebcf}$
\end{itemize}
while the remaining ones can be reduced to what we've already computed previously, i.e. 
\begin{equation}
\label{eqn:O2abL2cd}
    \avg{ O^{(2)}_{ab}(\bk) | L^{(2)}_{cd}(\bk') }' = - \left( \frac{k_a k_b}{k^2} - \frac{1}{3} \delta_{ab} \right) \left( \frac{k_c k_d}{k^2} - \frac{1}{3} \delta_{cd} \right) A_{O^{(2)}}(k), 
\end{equation}
where $A_{O^{(2)}}$ have been previously derived in the correlators of eqns. \ref{eqn:s2psi2}, \ref{eqn:dspsi2} and \ref{eqn:L2psi2}. Note that these latter correlators have only $m=0$ components prior to advection. \par 
The entire set of helicity spectra for shape--shape correlators at one loop is shown in Figure~\ref{fig:shapeshapeplot}. Each row shows the $m=\{0, 1, 2\}$ helicity spectra separated by columns of shape operator. The first column shows shape--shape correlators with dependence on the linear tidal field $s_{ab}$, the second those with dependence on the quadratic term $\delta s_{ab}$ and finally those with dependence on $s_{ac}s_{cb}$. The $(\delta s_{ab},\delta s_{ab})$ seem to have the highest amplitude for higher helicity spectra, and indeed we will see in \S\ref{sec:nbody} that this is also the most important contribution to the higher helicity spectra in the so-called local Lagrangian bias ansatz \cite{Akitsu23}. We will return to this point shortly when we compare our results to $N$-body simulations.

\section{Comparison to N-body}
\label{sec:nbody}

\begin{figure}
    \includegraphics[width=\textwidth]{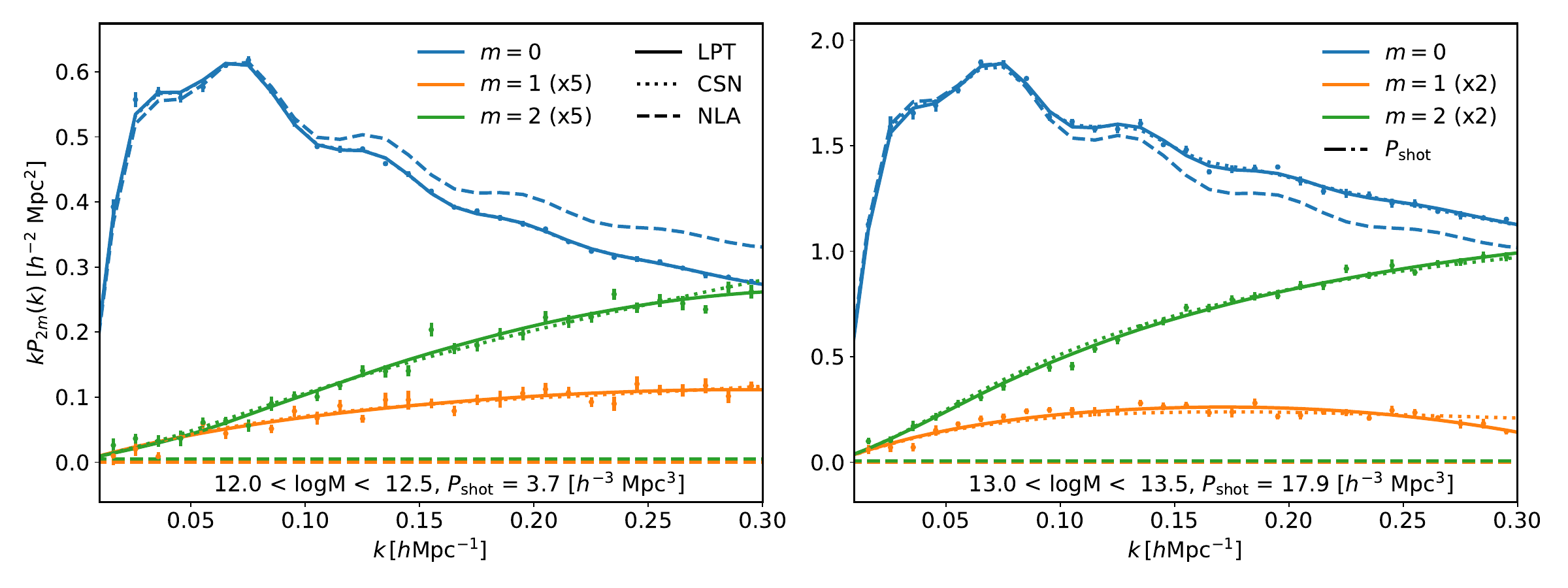}
    \caption{Perturbation theory fits to halo shape spectra of helicities $m=0, 1, 2$ with shape noise (SN) subtracted for clarity. The higher helicity spectra are suppressed relative to $m=0$ and are multiplied by constant factors to enhance their visibility. Solid and dotted lines show the results when using the full Lagrangian effective model with (LPT) and without (CSN) scale-dependent shape noise, while the dashed line shows the fit to the (non-)linear alignment model, also known as NLA. For lower mass halos both LPT and CSN represent excellent fits with reduced $\chi^2$ around one, while the scale-dependent shape noise is required to obtain a good fit for more massive halos. The NLA model fails to fit the $m=0$ spectrum past $k=0.1 \ihmpc$ and predicts zero for the higher-helicity spectra.}
    \label{fig:fits_to_data_NLA}
\end{figure}
In order to validate our perturbation theory we compare its predictions to the clustering of the three-dimensional shapes of halos measured in N-body simulations. In particular we will use the helicity auto spectra measured in ref.~\cite{Akitsu23}, which were taken from 8 boxes each with volume $(1.5 h^{-1} \text{Gpc})^3$. Unlike the projected $E$ and $B$ mode statistics used in previous works validating perturbation theory predictions for galaxy shapes, these spectra fully describe the three-dimensional clustering without mixing independent components, and moreover do not contain projection effects which break the isotropy of galaxy shapes in real space \cite{Bakx23}. Due to the orthogonality of the helicity components the $m = 0, 1, 2$ spectra can be treated as diagonal, and we will assume their covariance can be approximated as Gaussian.

The left and right panels of Figure~\ref{fig:fits_to_data_NLA} show the best fits of the theory to halos in the mass bins $12.0 < \log_{10} (M h\, /M_\odot) < 12.5$ and $13.0 < \log_{10} (M h\, /M_\odot) < 13.5$ at redshift $z = 0.51$, along with the shape-noise subtracted helicity spectra from N-body simulations. We also show a comparison with the standard (non)-linear alignment model adopted by stage-III cosmic shear surveys such as KiDS-1000~\cite{KiDS}. The NLA model is constructed by taking the linear Eulerian prediction for $(s_{ab}, s_{cd})$ and mapping the power spectrum to the non-linear power spectrum using a fitting function such as Halofit~\cite{Takahashi_2012}.

Lagrangian perturbation theory at 1-loop order, with all the independent bias and effective-theory parameters free, provides an excellent fit to all three helicity spectra for both the high and low mass N-body halos up to $k = 0.3 h$ Mpc$^{-1}$, as shown by the solid lines, with a reduced $\chi^2$ in both cases of approximately one. We note that these fits are performed assuming the full statistical power of the N-body volume ($27 h^{-3}$ Gpc$^3$). The NLA model, shown in dashed lines, on the other hand, cannot fit the $m=0$ spectrum past scales of $\sim 0.1 \ihmpc$ despite using the fully nonlinear matter power spectrum. Additionally, since the fiducial NLA modelling of shear surveys uses a fixed value for the shape noise (assumed to be the Poisson value) that is subtracted from the spectrum, with no additional freedom, the NLA model wholly fails to capture the higher helicity spectra -- which have been detected to be non-zero at high significance even after shot noise subtraction. 

The dotted lines show the prediction of perturbation theory without the scale-dependent ($k^2$) stochastic term. This term is rather non-negligible for the high mass bin, particularly in the $m=1, 2$ spectra which are 1-loop at leading order, and setting it to zero in our fits incurs a $\chi^2$ penalty mostly from these two curves. At first sight this may be surprising, given that halo \textit{densities} in the same mass bins are well-described assuming this term is negligible \cite{Chen20,Schmittfull21}. However, the importance of this term can be understood by noting that the deterministic bias contributions for shapes are suppressed by more than an order of magnitude compared  to those of densities, such that the leading (constant) stochastic contribution ($\propto R_h^3$) is a much larger fraction of the overall signal, as shown by the black dot-dashed line. Once this is taken into account, the scale dependence of the stochastic contribution ($\propto R_h^5 k^2$) is then roughly of the size preferred by the data. It is worth noting again that while there are three helicity spectra there are only two independent $k^2$ terms, which are enough to fit the data. Nonetheless, the prominence of the scale-dependent contribution for the $m= 1, 2$ spectra suggests that shape spectra observed in galaxy surveys, particularly for the $B$ modes, cannot be perturbatively modeled well past what is shown here even at higher orders in perturbation theory. On the other hand, these B mode contributions are all extremely subdominant to the shape noise, making their signal relatively suppressed and hard to detect in realistic surveys. 

\begin{figure}
    \includegraphics[width=\textwidth]{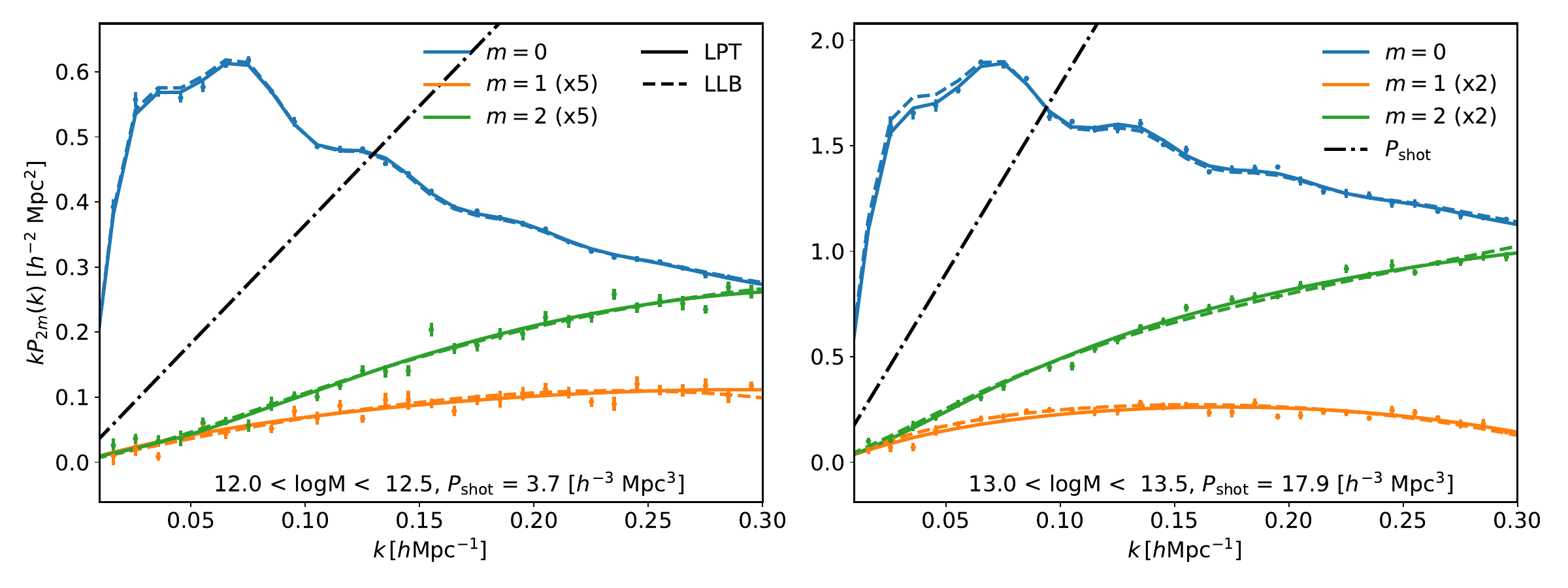}
    \caption{Fits to halo helicity spectra, as in Fig.~\ref{fig:fits_to_data_NLA}. The dashed lines now show the `local Lagrangian bias' model fit to these catalogs. The $\chi^2$ for LLB is approximately \edit{19 (55)}\% higher for the low (high) mass bins compared to the full LPT fit.}
    \label{fig:fits_to_data_llb}
\end{figure}

The above discussion centered on fits to galaxy shape spectra involving all free parameters at 1-loop in perturbation theory. Recently, ref.~\cite{Akitsu23} showed that halos are, at least approximately, well described as Lagrangian tracers of the initial shear field, up to multiplication by the (scalar) density. This suggests that an approximate biasing scheme for halos can be written as
\begin{equation}
    M_{ab}(\bq) = \left( 1 + b_1 \delta(\bq) + \frac12 b_2 (\delta^2(\bq) - \avg{\delta^2}) + b_s (s^2(\bq) - \avg{s^2}) + ...  \right)\ c_s s_{ab}(\bq).
\end{equation}
Since the cubic operators in this product are all degenerate with the linear bias $c_s$ this leads to a parameterization of the shape field with only two free dimensionless bias operators: $s_{ab}$ and $\delta s_{ab}$. The dashed lines in Fig.~\ref{fig:fits_to_data_llb} show the fits using this reduced parameter set, along with effective-theory contributions, when compared to the full LPT model described above. Evidently the fit is not perfect and we are unable to fully recover the shape power spectra on large scales; on the other hand, the resulting difference with N-body data is only at the $2\%$ or smaller level in terms of the linear bias. \edit{The increase in $\chi^2$ from this local Lagrangian bias (LLB) model is 
$\sim$ 19 (55) \% for the lower (higher) mass bins, respectively.
% $\sim$ 5 (17) \% for the lower (higher) mass bins, respectively.
The shifts in $\chi^2$ are higher than the change in degrees of freedom from restricting the parameter space of the alignment model. In addition, we note that the our fits to the full model are consistent with a Lagrangian bias model wherein only the quadratic terms ($s_{ab}, \delta s_{ab}, s^2_{ab}$, $L^{(2)}_{ab}$), in addition to the EFT corrections, are nonzero. This result is in good agreement with ref.~\cite{Akitsu23}, and in App.~\ref{app:mcmc} we report the results of two MCMC explorations of the parameter space with the same likelihood which we maximized in this section. We obtain highly significant detections of non-zero Lagrangian Zel'dovich terms, as well as the stochastic terms, for both halo mass bins. We also note a significant degeneracy in the cubic biases for ($L^{(3)}_{ab}$, $\delta t_{ab}$), which implies further dimensional reduction could be achieved for practical analyses. We also note that restricting the cubic bias terms to zero necessitates a non-zero value of the $L^{(2)}$ coefficient in order to obtain a comparable $\chi^2$, despite its posterior being consistent with zero in the full parameter exploration. We intend to return to this topic of model degeneracies in the near term with a more comprehensive set of simulations, halo samples, and redshifts to fully explore the degeneracies of our model and compare to previous measurements of shape parameters such as was done in ref.~\cite{Akitsu23}. Notably, the inclusion of the $m=0$ shape-matter cross-spectrum should help sharpen constraints on parameters from power spectrum-only fits \cite{Bakx23}.}

Nevertheless, since the intrinsic alignment signal is a small contaminant in cosmic shear data for source/lens redshift distributions that do not possess significant overlap, it seems likely that a reduced bias parameterization like LLB or Zeldovich-only can be used in lieu of the full model, assuming that the bias relations of halos continue to hold for galaxies. The net result would be a reduced parametrization for the galaxy shapes that is nonetheless more accurate than the TATT or NLA models alone. Analyses with such a reduced model could have implications for even current cosmic shear analyses~\cite{Secco_2022}, where degeneracies between intrinsic alignment model parameters (both bias coefficient amplitudes and their redshift-dependence) currently limit constraining power.

\section{Conclusions}
\label{sec:conclusions}

The cosmological distribution of galaxy shapes, known as intrinsic alignments, play an important role in current and upcoming galaxy surveys, both as one of the most significant contaminants in cosmic shear surveys and as a probe of large scale structure in their own right. In recent years, significant progress has been made towards the modeling these large-scale shape correlations, particularly within the language of perturbation theory, leading to consecutive development of the (non)linear alignment (NLA), quadratic tidal alignment and tidal torquing (TATT) and finally effective field theory (EFT) models \cite{hirataseljak04,Schmitz_2018,Vlah:2019byq}. The EFT model, in particular, represents a consistent treatment of perturbative contributions to intrinsic alignments up to a given order in the perturbation theory, and was shown in ref.~\cite{Bakx23} to be in excellent agreement with the shapes of halos in N-body simulations.

In this work we have continued this program, particularly as set out in ref.~\cite{Vlah:2019byq}, by formulating a description of intrinsic alignments within Lagrangian perturbation theory (LPT). Within this description the dynamics of galaxies are computed in terms of their displacements $\Psi$ from initial Lagrangian positions $\bq$, and their shapes described by a bias expansion in terms of their local initial conditions written as spatial derivatives of these displacements. This advective description of large-scale structure allows for a natural way to understand the large-scale displacements that damp the baryon acoustic oscillation (BAO) peak and other oscillatory features in the power spectrum, whose effect we resum in our calculations. We enumerate the contributions to this bias expansion up to cubic order and derive the 1-loop power spectrum of both galaxy shape--shape and galaxy density--shape correlations---the latter includes the special case of matter when all the Lagrangian bias coefficients are set to zero---representing a complete list of the relevant contributions to cosmic shear (`GI' and `II') and galaxy-galaxy lensing (`gI') measurements as well as the real-space contributions to spectroscopic shape analyses. At this order we require one linear, three quadratic and two cubic bias parameters. Our formalism also includes effective-theory counterterms and stochastic contributions necessary to control the impact of small-scale physics, including two next-to-leading order scale-dependent contributions to the latter. We make \texttt{spinosaurus}, a \texttt{Python} code that performs these calculations using Hankel transforms and \texttt{FFTLog}, publically available.

As a test of the theory model developed in this work, we fit our theory predictions to the three-dimensional helicity spectra of simulated halos measured in ref.~\cite{Akitsu23}. These fits capture the full angular structure of galaxy spin statistics at the 2-point function level and are distinct from previous comparisons in the literature which used only the statistics of projected E and B modes. Perturbation theory at 1-loop order gives an excellent fit to the shape-shape autospectra of these halos, with $\chi^2$ per degree of freedom close to unity, out to $k = 0.3 h$ Mpc$^{-1}$ at $z = 0.51$, and we find a strong preference for including a $k^2$ scale dependence to the shot noise, particularly in order to fit the $m = 2$ spectrum. We also test the ``local Lagrangian bias'' ansatz, previously seen to be a good approximation at the field \cite{Akitsu23} and projected 2-point function \cite{Bakx23,Maion23} levels. This ansatz involves only two free bias parameters in addition to the counter and stochastic terms---we find that it is able to reasonably describe the data in qualitative terms but results in a mis-estimation of the linear bias at the few percent level when fit over the same range of scales, leading to a non-negligible $\chi^2$ difference. Nonetheless, we note that our fits are performed for full three-dimensional measurements with total volume $27 h^{-3}$ Gpc$^3$, representing greater constraining power than near-term cosmic shear or spectroscopic shape surveys, such that any of the schemes we have introduced in this work would likely be acceptable in realistic applications despite the full model being the only one that adequately fits the simulations at their true statistical power.

Let us close by mentioning a few possible extensions to this work. Most directly, our model can be applied to analyses of existing weak lensing data sets for both cosmic shear alone and galaxy--galaxy lensing, though scale cuts would necessarily need to be applied in order to limit the contributions of intrinsic alignments to these data to perturbative scales. Recently, ref.~\cite{Maion23} proposed a hybrid model \cite{Modi_2020} of intrinsic alignments wherein Lagrangian bias operators are advected by the non-perturbative displacements of dark matter solved for in N-body simulations, with the aim of extending the reach of the perturbative bias expansion beyond the dynamical nonlinear scale. The success of such an approach is related to the ability of the limited Lagrangian model described above to qualitatively describe the data. Our calculations in this paper provide direct predictions for the components of the hybrid model and will be essential in producing a viable emulator for it to interpolate across different cosmologies in data analyses. Our calculations similarly serve as building blocks for extensions of the technique of `Zel'dovich control variates'~\cite{Kokron_ZCV, DeRose_ZCV,hadzhiyska2023mitigating} to the realm of intrinsic alignment statistics -- allowing for more accurate and precise measurements of IA power spectra from $N$-body simulations.

Recently, there has been a renewed interest in measuring intrinsic alignments not just as a contaminant in lensing surveys but as a cosmological signal in their own right, particularly in the context of spectroscopic surveys. Perhaps the most exciting extension of this work is to the modeling of galaxy shape statistcs in redshift space, where these surveys operate. In such surveys anisotropies known as redshift-space distortions arise due to the degeneracy between cosmological redshifts and peculiar velocities along the line of sight, and manifest as a boost $\textbf{u} = \textbf{v}_\parallel / \mathcal{H}$ to the line-of-sight positions of galaxies. Within Lagrangian perturbation this amounts to boosting the order $n$ displacement by \cite{Matsubara08a,Carlson13}
\begin{equation}
    \Psi^{(n),s}_i(\bq) = \Psi^{(n)}_i(\bq) + \mathbf{u}_i^{(n)} = R^{(n)}_{ij} \Psi^{(n)}_j(\bq)
\end{equation}
where we have defined the matrix $R^{(n)}_{ij} = \delta_{ij} + n f(a) \hat{n}_i \hat{n}_j$, such that the redshift space power spectrum becomes
\begin{equation}
    P_{s,abcd}(\bk) = \int d^3\bq \ e^{i\bk\cdot\bq} \avg{  e^{i\bk\cdot\Delta_s}\ F_{ab}(\bq_1) F_{cd}(\bq_2)}.
\end{equation}
where $\Delta_{s,i}^{(n)} = R^{(n)}_{ij} \Delta^{(n)}_j$ is the pairwise distortion in redshift space. Calculating the power spectrum in redshift space is then equivalent to boosting the pairwise-displacement and bias operator correlators we have already computed at each order $n$ with the matrix $R^{(n)}_{ij}$, though we caution that the resulting angular integrals will be slightly more complicated than in this work \cite{Vlah19,Chen_2021} -- the resulting power spectra will now only be symmetric about rotations about the line-of-sight and thus require a different tensor basis \cite{Matsubara22a,Matsubara22b,Matsubara22c}. Alternatively the displacements can be perturbatively expanded to yield a moment expansion in the velocities, also known as the distribution function approach to redshift-space distortions \cite{Seljak11,Vlah19,Chen20}. In this case we can write
\begin{equation}
    P_{s,abcd}(\bk) = \sum_{n=0}^\infty \frac{i^n}{n!} k_{i_1} ... k_{i_n}  \int d^3\bq\ e^{i\bk\cdot\bq} \avg{e^{i\bk\cdot\Delta} \Delta \textbf{u}_{i_1} ... \Delta \textbf{u}_{i_n} F_{ab}(\bq_1) F_{cd}(\bq_2)}
\end{equation}
where again $\Delta \textbf{u}$ are simply scalar multiples of $\Delta$ given by their time derivatives, precisely what we have painstakingly computed in this paper to 1-loop order.  We intend to return to this topic in future work.

\acknowledgments
We thank Joe DeRose, Zvonimir Vlah, Martin White for useful discussions and Elisabeth Krause for helpful comments on a draft of this work. We especially thank Kazuyuki Akitsu for frequent conversations throughout the preparation of this work that greatly aided our understanding of the subject, as well as for providing his measurements of halo shape spectra. S.C. is supported by the Bezos Membership at the Institute for Advanced Study. N.K. would like to thank the International Center for Theoretical Physics -- South American Institute for Fundamental Research (ICTP-SAIFR) for their hospitality during part of the completion of this work. ICTP-SAIFR is funded by FAPESP grant 2021/14335-0. Figures and code to produce them in this work have been made using the SciPy Stack \citep{2020NumPy-Array,2020SciPy-NMeth,4160265} and \texttt{GetDist}~\cite{lewis2019getdist}. This research has made use of NASA's Astrophysics Data System and the arXiv preprint server.

\appendix

\section{Tensor analogs of Legendre polynomials}
\label{app:tensorlegendre}
We collect definitions of the tensor Legendre polynomials used throughout the text, up to fourth order. The fifth order polynomial contains many terms and we refer to the included Mathematica notebooks for its presentation and use.
\begin{align}
    &P_0 (\hat{k}) = 1 \\
    &P_1^i (\hat{k}) = \hat{k}_i \\
    \label{eqn:p2legendre} &P_2^{ij} (\hat{k}) = \frac{3}{2} \left ( \hat{k}_i \hat{k}_j - \frac{1}{3} \delta_{ij} \right )\\ 
    &P_3^{iab} (\hat{k}) =  \frac{1}{2} \left ( 5 \hat{k}_i\hat{k}_a \hat{k}_b - (\hat{k}_i \delta_{ab} + \hat{k}_{(a}\delta_{b)i} \right ) \\
    &P_{4}^{ijab} (\hat{k}) = \frac{1}{8} \left (35 \hat{k}_i \hat{k}_j \hat{k}_a \hat{k}_b - 5 ( \hat{k}_i\hat{k}_j\delta_{ab} + \hat{k}_a\hat{k}_b \delta_{ij} + \hat{k}_i\hat{k}_{(a}\delta_{b)j}+ \hat{k}_j \hat{k}_{(a}\delta_{b)i}) + (\delta_{ij}\delta_{ab} + \delta_{i(a}\delta_{b)j}) \right )
\end{align}

\section{Order-by-order contributions to derivatives of the generating function at 1-loop}
\label{app:cumulant_contributions}
In this appendix, we enumerate contributions to the generating function, or its derivatives, to 1-loop order in perturbation theory.

\subsection{Matter Tracer Cross Spectrum}
In the case of the matter-tracer cross spectrum we need to look for terms with only one power of $\alpha$, i.e.
\begin{equation*}
\left. \frac{\partial M}{\partial \alpha_2^a} \right |_{\boldsymbol{\alpha}=0} = M(\mathbf{0}) \Big( i k_i \langle \Delta_i O_2^a \rangle - \frac{1}{2} k_i k_j \langle \Delta_i \Delta_j O_2^a \rangle + ... \Big)
\end{equation*}
The contributions are
\begin{align}
 &\langle \Delta_i O_2^a \rangle \ni \langle \Delta_i^{(1)} O_2^{a,(1)} \rangle, \langle \Delta_i^{(2)} O_2^{a,(2)} \rangle, \langle \Delta_i^{(1)} O_2^{a,(3)} \rangle, \langle \Delta_i^{(3)} O_2^{a,(1)} \rangle \nonumber \\
 &\langle \Delta_i \Delta_j O_2^a \rangle \ni \langle \Delta_i^{(2)} \Delta_j^{(1)} O_2^{a,(1)} \rangle, \langle \Delta_i^{(1)} \Delta_j^{(1)} O_2^{a,(2)} \rangle
\end{align}

\subsection{Tracer-Tracer Spectrum}
For the tracer--tracer spectrum two derivatives are needed
\begin{equation*}
\left. \frac{\partial^2 M}{\partial \alpha_1^a \partial \alpha_2^b}\right |_{\boldsymbol{\alpha}=0} = M(\mathbf{0}) \Big( -k_i k_j \langle \Delta_i O_1^a \rangle\langle \Delta_j O_2^b \rangle + \langle O_1^a O_2^b \rangle + ik_i \langle \Delta_i O_1^a O_2^b \rangle_c + ... \Big)
\end{equation*}
Again we can enumerate the relevant contributions by order:
\begin{align}
    &\langle \Delta_i O_1^a \rangle\langle \Delta_j O_2^b \rangle \ni \langle \Delta_i^{(1)} O_1^{a,(1)} \rangle\langle \Delta_j^{(1)} O_2^{b,(1)} \rangle \nonumber \\
    &\langle O_1^a O_2^b \rangle \ni \langle O_1^{a,(1)} O_2^{b,(1)} \rangle, \langle O_1^{a,(1)} O_2^{b,(3)} \rangle, \langle O_1^{a,(3)} O_2^{b,(1)} \rangle,  \langle O_1^{a,(2)} O_2^{b,(2)} \rangle \nonumber \\
    &\langle \Delta_i O_1^a O_2^b \rangle_c \ni \langle \Delta_i^{(1)} O_1^{a,(1)} O_2^{b,(2)} \rangle_c, \langle \Delta_i^{(1)} O_1^{a,(2)} O_2^{b,(1)} \rangle_c, \langle \Delta_i^{(2)} O_1^{a,(1)} O_2^{b,(1)} \rangle_c
\end{align}

\section{Definitions of LPT correlators}
\label{app:lagcorrs}
In this section we review the Lagrangian correlators defined in refs.~\cite{White_2014,Kokron22}. Specifically we will list the correlators including at least one power of the Lagrangian tidal field $s_{ab}(\bq)$. Readers interested in the remaining correlators (e.g.$\ A_{ij}, U_i, \xi_L$ etc.) are directed to ref.~\cite{Carlson13}. The first correlator as discussed in \S\ref{ssec:angular} is
\begin{equation}
    E_{ab}(\bq) = \avg{\delta(\bq_1) s_{ab}(\bq_2)} = \mathcal{J}_1(q) (\delta_{ab}/3 - \hq_a \hq_b) = -\frac{2}{3} \xi^2_0(q) P_{2,ab}(\hq).
\end{equation}
where $\mathcal{J}_1 = \xi^2_0$.  Ref.~\cite{White_2014} defined a series of functions $\mathcal{J}_n$ via sums of generalized correlation functions, and we will follows that convention here. The next is the shear-displacement correlator
\begin{equation*}
    B_{iab}(\bq) = \avg{\Delta s_{ab}(\bq_2)} = \delta_{ab} \hq_i \mathcal{J}_2 + (\delta_{ia} \hq_b + \delta_{ib} \hq_a) \mathcal{J}_3 + \hq_i \hq_a \hq_b \mathcal{J}_4
\end{equation*}
where
\begin{equation}
    \mathcal{J}_2 = \frac{2}{15} \xi^1_{-1} - \frac15 \xi^3_{-1},\ \mathcal{J}_3 = -\frac15 \xi^1_{-1} - \frac15 \xi^3_{-1},\ \mathcal{J}_4 = \xi^3_{-1}.
\end{equation}
As discussed in \S\ref{ssec:angular} this correlator can equivalently be written in the Legendre basis adopted in this work as\begin{equation}
    B_{iab}(\bq) = \frac{2}{5} \xi^{3}_{-1}(q) \mathcal{Q}^3_{iab}(\hq) + \frac{2}{15} \xi^{1}_{-1}(q)\mathcal{Q}^1_{iab}(\hq).
\end{equation}
Finally the shear-shear correlator is
\begin{align*}
    C_{abcd}(\bq) = \avg{s_{ab}(\bq_1) s_{cd}(\bq_2) } =\  &\delta_{ab} \delta_{cd} \mathcal{J}_5 + (\delta_{ac} \delta_{bd} + \delta_{ad} \delta_{bc}) \mathcal{J}_6 \\
    &+ (\delta_{ab} \hq_c \hq_d + \delta_{cd} \hq_a \hq_b) \mathcal{J}_7 \\
    &+ (\delta_{ac} \hq_b \hq_d + \delta_{bd} \hq_a \hq_c + \delta_{ad} \hq_b \hq_c + \delta_{bc} \hq_a \hq_d) \mathcal{J}_8 \\
    &+ \hq_a \hq_b \hq_c \hq_d \mathcal{J}_9
\end{align*}
where
\begin{equation*}
    \mathcal{J}_5 = -\frac{1}{315} ( 14 \xi^0_0 + 40 \xi^2_0 - 9 \xi^4_0),\ \mathcal{J}_6 = \frac{1}{105} (7 \xi^0_0 + 10 \xi^2_0 + 3 \xi^4_0)
\end{equation*}
\begin{equation}
    \mathcal{J}_7 = \frac{1}{21} (  4 \xi^2_0 - 3 \xi^4_0),\ \mathcal{J}_8 = -\frac{1}{7} ( \xi^2_0 + \xi^4_0),\ \mathcal{J}_9 = \xi^4_0.
\end{equation}
Again we can equivalently express this correlator in terms of the Legendre basis in which
\begin{equation}
    C_{abcd}(\bq) = \frac{8}{35} \xi^4_0(q) \mathcal{Q}^4_{abcd}(\hq) + \frac{8}{63} \xi^2_0(q) \mathcal{Q}^2_{abcd}(\hq) - \frac{2}{45} \xi^0_0(q) \mathcal{Q}^0_{abcd}(\hq).
\end{equation}

\section{Explicit Expressions for Loop Integrals}
\label{app:integral_details}

\subsection{ $\avg{s_{1,ab} \Delta^{(1)}_i \Delta^{(2)}_j}$ }
For the first 22-type term we can write
\begin{align}
\label{eqn:quaddeltasab}
    \langle (\Psi^{(1)}_i s_{ab})(\bk)\; | \Psi^{(2)}_j(\bk') \rangle' &= -  \frac17 \frac{i k_j}{k^2} \langle (\Psi^{(1)}_i s_{ab})(\bk)\; |\;  \big(\delta^2 - \frac{3}{2} s^2 \big)(\bk') \rangle' \nonumber \\
    &= - \frac{1}{7} \frac{i k_j}{k^2} \text{FT}\Big\{ 2 U_i(\bq) E_{ab}(\bq) -3 B_{icd}(\bq) C_{abcd}(\bq) \Big\} \nonumber \\
    &= - \frac{i k_j}{k^2} \text{FT}\Big\{ \tilde{F}^s_3(q) \mathcal{Q}^3_{iab}(\hq) +  \tilde{F}^s_1(q) \mathcal{Q}^1_{iab}(\hq) \Big\} \nonumber \\
    &\equiv \frac{k_j}{k^2} \Big( Q_3^{s}(k) P_3(\hk) - Q_1^s(k) (\hk_i \delta_{ab} - \frac{3}{2} \hk_{(a} \delta_{b)i}) \Big)
\end{align}
where $\tilde{F}^s_{1,3}$ can be obtained by projecting the correlators in curly brackets in the second line into the $\mathcal{Q}(\hq)$ basis, and $Q^s_{1,3} = \xi^{1,3}_0[\tilde{F}^s_{1,3}]$.

The integrals for the 13-type term can similarly be directly projected onto the $\mathcal{Q}(\hk)$ basis such that the integrals can be rewritten in terms of generalized correlation functions:
\begin{align}
    R_3^{s}(k) = \frac{2}{5} P(k) \int_0^\infty dq\ q\   &\Big[-\frac{2}{35} k \xi^0_0(q) j_0(kq)+\frac{18}{245} \xi^1_1(q) j_1(kq)+\frac{10}{49} k \xi^2_0(q) j_2(kq) \nonumber  \\ 
    &-\frac{22}{105} \xi^3_1(q) j_3(kq) -\frac{36}{245} k \xi^4_0(q) j_4(kq)+\frac{20}{147} \xi^5_1(q) j_5(kq) \Big] \\
    R_1^{s}(k) = \frac{1}{15}P(k) \int_0^\infty dq\ q\ & \Big[ \frac{4}{35} k \xi^0_0(q) j_0(kq)+\frac{12}{35} \xi^1_1(q)j_1(kq)-\frac{20}{49} k \xi^2_0(q) j_2(kq) -\frac{12}{35} \xi^3_1(q) j_3(kq) \nonumber \\
    &  +\frac{72}{245} k \xi^4_0 (q) j_4(kq)  \Big].
\end{align}

\subsection{$\avg{O^{(2)}_{ab} \Delta^{(1)}_i \Delta^{(1)}_j}$}
Again pulling out a factor of $i k^{-1} \hk$ we can write
\begin{equation}
    \avg{\Psi^{(1)}_i L^{(2)}_{ab}(\bk) | \Psi^{(1)}_j(\bk')} = \frac{k_j}{k^2}\left[R_3^{L^{(2)}}(k) \mathcal{Q}^3_{iab}(\hk) + R_1^{L^{(2)}}(k) \mathcal{Q}^1_{iab}(\hk) \right]
\end{equation}
where we have defined the kernels
\begin{align}
    R_3^{L^{(2)}}(k) &= -P(k)\int dq\ q\ \Big[ \frac{12}{175} k^2 \xi^1_{-1}(q) j_1(kq) -\frac{12}{175} k^2 \xi^3_{-1}(q) j_3(kq)+\frac{8}{175} k \xi^0_0(q) j_0(kq)
   j_2(kq) \nonumber \\
   &-\frac{8}{49} k \xi^2_0(q)  +\frac{144 k \xi^4_0(q) j_4(kq)}{1225}-\frac{36 \xi^1_1(q) j_1(kq)}{1225}+\frac{44}{525} \xi^3_1(q) j_3(kq)-\frac{8}{147} \xi^5_1(q) j_5(kq) \Big] \nonumber
\end{align}
\begin{align}
    R_1^{L^{(2)}}(k) = -P(k)\int& dq\ q\ \Big[ -\frac{4}{175} k^2 \xi^1_{-1}(q) j_1(kq) + \frac{4}{175}k^2 \xi^3_{-1}(q) j_3(kq)+\frac{32}{525} k \xi^0_0(q) j_0(kq)  \nonumber \\
   &  -\frac{8}{147} k \xi^2_0(q)
   j_2(kq)-\frac{8 k \xi^4_0(q) j_4(kq)}{1225}-\frac{4}{175} \xi^1_1(q) j_1(kq)+\frac{4}{175} \xi^3_1(q) j_3(kq) \Big]. \nonumber
\end{align}
\section{Parameter posteriors from MCMC}
\label{app:mcmc}
\begin{figure}
    \centering
    \includegraphics[width=\textwidth]{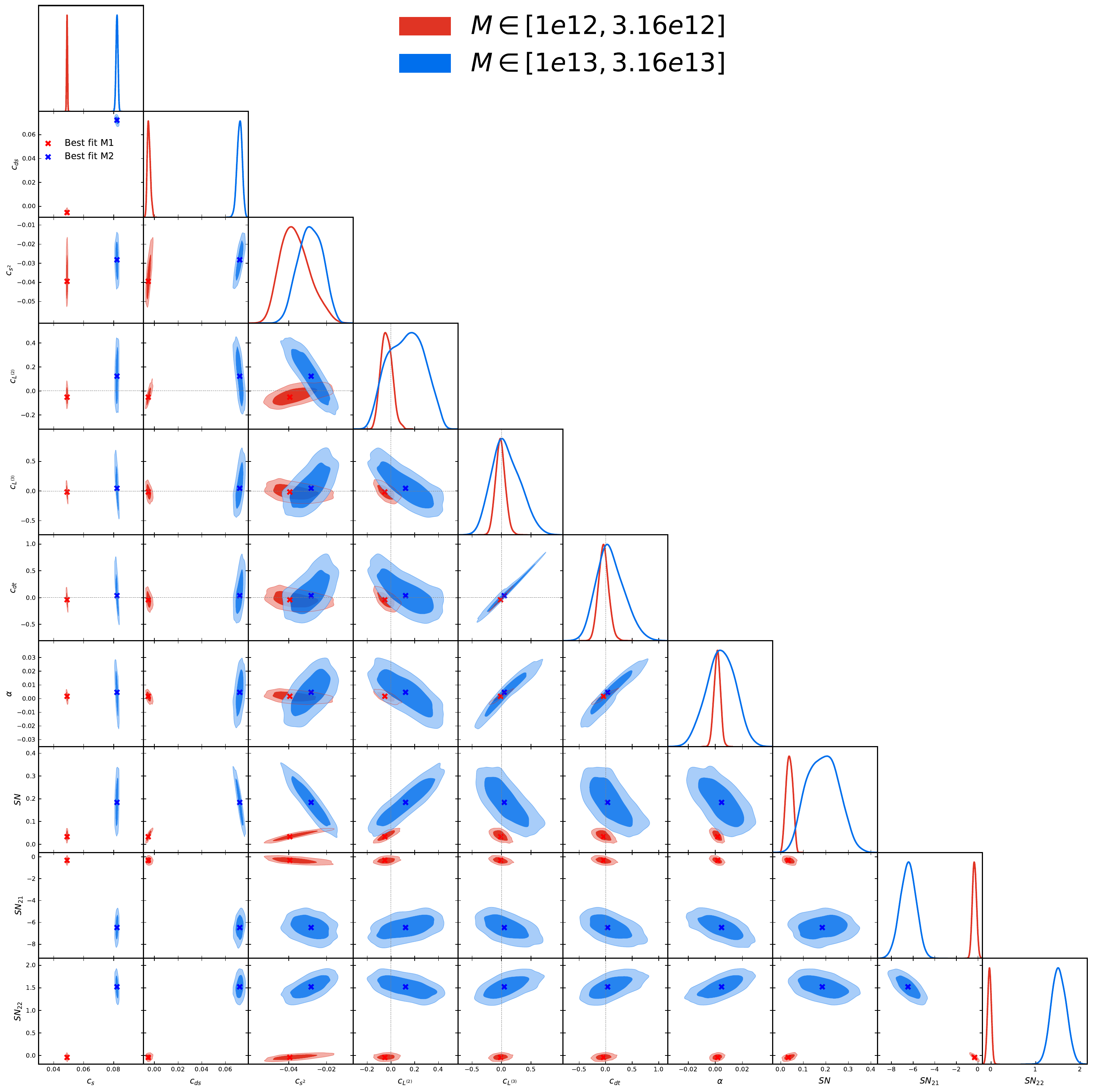}
    \caption{Results from exploring the full model parameter space using MCMC for the low and high halo mass samples we fit in the main text. The best-fit values for each sample are marked by the colored `x'. For the higher-order bias parameters we add markers at zero to guide the eye.}
    \label{fig:mcmc}
\end{figure}

In Fig.~\ref{fig:mcmc} we report the resulting model posteriors from fitting the full IA model to the halo samples discussed in Sec.~\ref{sec:nbody}. The higher stochasticity in the high mass sample leads to a significant degradation in the constraints of higher order IA parameters, although we note that $c_{s^2}$ is still constrained to be non-zero at high significance. We also report  a mild preference for non-zero $c_{L^{(2)}}$ which is required to ensure a good $\chi^2$. We sample 4 independent chains using \texttt{emcee}~\cite{2013PASP..125..306F} with $N=20$ walkers, until their Gelman-Rubin statistic~\cite{gelmanrubin} reaches $R-1 \leq 0.01$, at which point we consider them converged. 
\bibliography{refs}
\bibliographystyle{jhep}
\end{document}